\documentclass[manuscript,screen]{acmart}

\usepackage[utf8]{inputenc}
\usepackage{longtable} 
\usepackage{tikz}
\usepackage{amsfonts}
\usepackage{amsmath}
\usepackage[most]{tcolorbox}
\usepackage{graphicx}
\usepackage{booktabs}
\usepackage{array}
\usepackage{geometry}
\usepackage{tabularx}
\usepackage{subcaption}
\usepackage{booktabs} 
\usepackage{mdframed}

\usepackage{multirow}
\usepackage{lipsum} 

\definecolor{bluered}{RGB}{0, 0, 0}

\newcommand{\sq}[1]{`#1'}
\newcommand{\dq}[1]{``#1''}

\tcbset{
    examplebox/.style={
        enhanced jigsaw, 
        colback=white,
        boxrule=0.5pt,
        colframe=black!15,
        fonttitle=\bfseries,
        coltitle=black,
        colbacktitle=white,
        toptitle=1mm,
        bottomtitle=1mm,
        left=2mm,
        right=2mm,
        before skip=10pt plus 2pt,
        after skip=10pt plus 2pt,
        sharp corners,
        boxsep=0pt,
        left=8pt,
        right=8pt,
        top=4pt,
        bottom=4pt,
        title style={left=8pt},
        separator sign none
    },
    titlebox/.style={
        colback=gray!20,
        colframe=gray!20,
        fonttitle=\bfseries,
        coltitle=black,
        left=2mm,
        right=2mm,
        boxsep=0pt,
        left=8pt,
        right=8pt,
        top=4pt,
        bottom=4pt
    }
}

\definecolor{bluegreen}{RGB}{0, 0, 0}

\newcommand{\bluegreen}[1]{{\color{bluegreen}#1}}

\definecolor{hypothesisColor}{RGB}{220,230,240} 
\definecolor{conclusionColor}{RGB}{210,250,210} 
\usepackage{colortbl} 

\AtBeginDocument{%
  \providecommand\BibTeX{{%
    \normalfont B\kern-0.5em{\scshape i\kern-0.25em b}\kern-0.8em\TeX}}}

\setcopyright{acmcopyright}
\copyrightyear{2024}
\acmYear{2024}
\acmDOI{XXXXXXX.XXXXXXX}

\begin{document}



\title{Understanding Biases in ChatGPT-based Recommender Systems: Provider Fairness, Temporal Stability, and Recency}


\author{Yashar Deldjoo}
\email{deldjooy@acm.org}
\orcid{0000-0002-6767-358X}
\affiliation{%
  \institution{Polytechnic University of Bari}
  \streetaddress{P.O. Box 1212}
  \city{Bari}
  \state{Via Orabana, 4}
  \country{Italy}
  \postcode{70125}
}




\renewcommand{\shortauthors}{Yashar Deldjoo}

\begin{abstract}

This paper explores the biases inherent in ChatGPT-based recommender systems, focusing on provider fairness (item-side fairness). Through extensive experiments and over a thousand API calls, we investigate the impact of prompt design strategies—including structure, system role, and intent—on evaluation metrics such as provider fairness, catalog coverage, temporal stability, and recency. The first experiment examines these strategies in \textbf{classical top-K recommendations,} while the second evaluates \textbf{sequential in-context learning (ICL}).

In the first experiment, we assess seven distinct prompt scenarios on top-K recommendation accuracy and fairness. Accuracy-oriented prompts, like Simple and Chain-of-Thought (COT), outperform diversification prompts, which, despite enhancing temporal freshness, reduce accuracy by up to 50\%. Embedding fairness into system roles, such as \dq{act as a fair recommender,} proved more effective than fairness directives within prompts. We also found that diversification prompts led to recommending newer movies, offering broader genre distribution compared to traditional collaborative filtering (CF) models. The system showed high consistency across multiple runs.

The second experiment explores sequential ICL, comparing zero-shot and few-shot learning scenarios. Results indicate that including user demographic information in prompts affects model biases and stereotypes. However, ICL did not consistently improve item fairness and catalog coverage over zero-shot learning. Zero-shot learning achieved higher NDCG and coverage, while ICL-2 showed slight improvements in hit rate (HR) when age-group context was included. Overall, our study provides insights into biases of RecLLMs, particularly in provider fairness and catalog coverage. By examining prompt design, learning strategies, and system roles, we highlight the potential and challenges of integrating large language models into recommendation systems, paving the way for future research. Further details can be found at \href{https://github.com/yasdel/Benchmark\_RecLLM\_Fairness}{https://github.com/yasdel/Benchmark\_RecLLM\_Fairness}.
\end{abstract}




\received{January 2024}

\maketitle

\section{Introduction}
\label{sec:intro}

\noindent \bluegreen{\textbf{Context.} Recommender systems are integral to various large-scale internet services, benefiting consumers, producers, and other stakeholders in multi-stakeholder markets~\cite{deldjoo2023fairness,abdollahpouri2021multistakeholder,ekstrand2019fairness,zheng2019multi}. The advent of generative models, particularly large language models (LLMs), holds the promise of offering better personalization experiences. LLMs enable conversational natural language (NL) interactions~\cite{he2023large,biancofiore2024interactive}, unlocking rich NL data like item descriptions, reviews, and queries within recommender systems (RS). Harnessing the general reasoning abilities of pretrained LLMs allows for addressing diverse, nuanced NL user preferences and feedback through highly personalized interactions. This contrasts with rigid ID-based methods that heavily rely on non-textual data.~\cite{deldjoo2024review,deldjoo2024recommendation}. This paper focuses on the emerging role of LLMs, specifically ChatGPT, in recommender systems and scrutinizes \textit{their biases,} with a particular emphasis on \dq{item-side fairness.} 

Item-side fairness is important to ensure diverse item groups receive fair exposure, benefiting item producers such as micro-businesses in job recommendations and promoting content related to vulnerable populations. Recent research shows that LLM-based recommender systems (RecLLMs), with their reliance on semantic clues and generative capabilities, can introduce unique biases not present in conventional systems~\cite{deldjoo2024cfairllm,jiang2024item,deldjoo2023fairness,deldjoo2024fairevalllm,zhang2023chatgpt}. Therefore, it is essential to investigate how these biases manifest and impact various stakeholders.}

\subsection{Background and Motivation}
\bluegreen{At a high level, the idea of generative modeling involves creating applications that not only make decisions based on data but also generate new data by learning patterns. This powerful idea has enabled various applications in AI disciplines, such as image generation, text synthesis, and music composition~\cite{min2023recent,koh2024generating,ferreira2023generating,zhang2023text}. Generative modeling has recently gained prominence due to advances in model paradigms such as Generative Adversarial Networks (GANs)~\cite{goodfellow2014generative}, Variational Autoencoders (VAEs)~\cite{kingma2013auto}, Diffusion Models~\cite{sohl2015deep,ho2020denoising}, and Transformer-based architectures such as GPT~\cite{wei2022emergent, sparks_of_agi} and other LLMs. In the recommender system (RS) field, generative models are not entirely new and have been used in various capacities for tasks such as data synthesis and augmentation~\cite{chae2019rating, yuan2020exploring}, model regularization \cite{liang2018variational, sachdeva2019sequential, ma2019learning}, and generating complex recommendation structures \cite{jiang2018beyond, liu2021variation}. These advancements fall under Deep Generative Models (DGMs), which combine traditional probabilistic models with Deep Neural Networks (DNNs). The core strength of DGMs lies in their ability to model and sample from the data distribution they are trained on for various inferential purposes.

\begin{figure}[!t]
    \centering
    \includegraphics[width =1.00\linewidth]{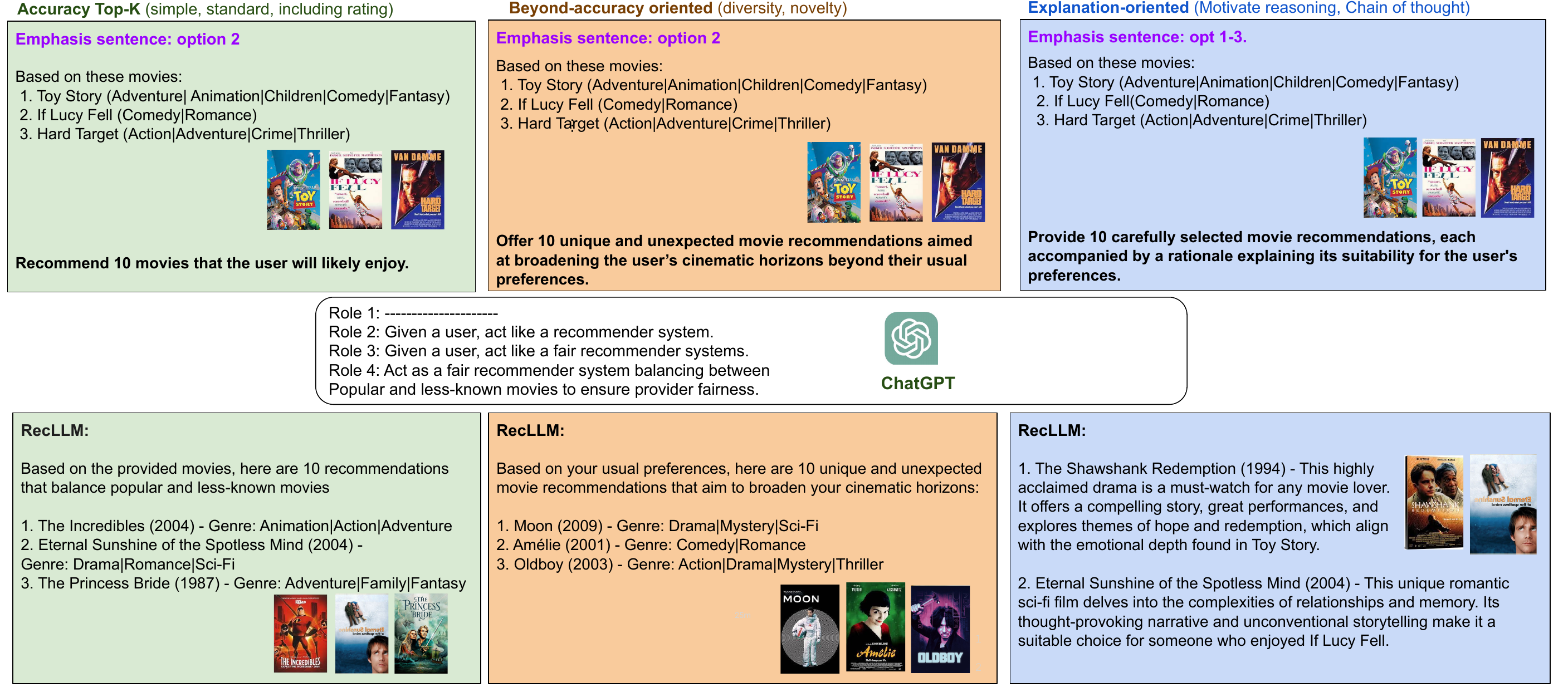}
    \caption{Conceptual idea behind experiment 1, prompt-design scenarios.} 

    \label{fig:conceptual}
\end{figure}

However, these models have experienced a resurgence of interest in the RS community, similar to trends in the broader AI landscape. This resurgence can be partly attributed to the introduction of LLMs such as ChatGPT, which have significantly improved the natural language understanding capabilities of these systems. LLMs demonstrate impressive in-context learning (ICL) and few-shot generalization abilities, making them valuable assets in both direct and indirect applications within recommender systems, where our goal here is to use them as direct recommenders (shown in Figure~\ref{fig:conceptual}). 


Notwithstanding their great success, the vast and unregulated nature of the internet data used to train LLMs raises concerns about possible biases against specific races, genders, popular brands, and other sensitive attributes that could be encoded in these networks. For example, if an LLM is predominantly trained on data from popular e-commerce sites, it might disproportionately recommend products from more recognized brands, overlooking niche or emerging brands. Similarly, biases in language around gender or race could skew recommendations in subtle but impactful ways. Another example could be the over-representation of content from certain geographical regions, leading to a lack of diversity in recommended media or news articles. Hence, the unchecked employment of these systems in commercial recommender systems may lead to unfair treatment of minority groups, reinforcing existing stereotypes, or exacerbating economic disparities.

This research explores different biases of ChatGPT-based RecLLMs, specifically on the \textbf{provider side,} to understand how these biases manifest in recommendations and affect various stakeholders. The primary goal of this research is two-fold:
\begin{enumerate}
    \item To enhance our understanding of the general performance of ChatGPT-based RecLLMs in terms of accuracy, provider fairness, and other nuanced evaluation dimensions (genre dominance, temporal stability, temporal freshness); and
\item To determine whether factors such as prompt design and ICL type can be leveraged in RecLLM design to enhance item fairness without compromising accuracy;
\end{enumerate}

\noindent To the best of our knowledge, the factors considered in this study are novel and have not been given enough attention in previous research. We believe that combining these studies into a single research effort opens up new avenues for future research in RecLLMs. Through \textbf{two} separate but extensive experiments, we introduce the following catalysts in the current work in hand:
\begin{itemize}
\item \textbf{Prompt Design.} We examine how different prompt structures and instructions affect recommendation outcomes. As shown in Figure~\ref{fig:conceptual} and Table~\ref{tab:prompt_scen}, we designed seven prompt scenarios classified into three categories: \textit{accuracy-oriented}, \textit{beyond-accuracy}, and \textit{reasoning}. For instance, prompts instructing the model to diversify resulted in recommendations that were temporally fresher but not significantly novel or diverse, leading to a significant reduction in accuracy (e.g., NDCG) by 50\%. In contrast, accuracy-oriented and reasoning-focused prompts, such as Simple and Chain-of-Thought (COT), performed much better.
    
    \item \textbf{System Roles.} We further investigated the impact of assigning specific \textbf{system roles}, such as \dq{acting as a fair recommender.} We observed that incorporating statements about fairness tends to improve system quality more effectively than including fairness directives directly within the prompt.

    \item \textbf{Stability Over Time.} We considered that external factors such as trending data or algorithmic updates might introduce randomness into the responses of GPT-based models. To study this, each experiment was conducted five times specifically for GPT-based models. We assessed the consistency of recommendations over multiple runs to ensure reliability.

    \item \textbf{In-Context Learning (ICL).} We analyzed the effect of zero-shot versus few-shot in-context learning (ICL) on provider fairness, focusing on demographic information impact on recommendations. We noted that incorporating user demographic data in prompts affected model biases and stereotypes. We structured the user profile into context, example, and demographic information parts. Our study highlights the nuanced effects of user profile attributes and sampling strategies on recommender systems.
\end{itemize}}
\vspace{2mm}
\noindent \textbf{Example related to Experiment 1.} We illustrate the impact of different \dq{prompt design} strategies on the output of ChatGPT-based recommendation systems through a clear example, as shown in Figure \ref{fig:conceptual}. We use a randomly chosen user profile from the MovieLens dataset to demonstrate this effect. The user's historical interactions with the system include a diverse range of movies such as \dq{Toy Story} (categorized with genres  Adventure|Animation|Children|Comedy|Fantasy), \dq{If Lucy Fell} (Comedy|Romance), and \dq{Hard Target} (Action|Adventure|Crime|Thriller).

To highlight the differences in recommendations based on the design of the system, we designed a total of \underline{seven} prompts categorized into \underline{three} distinct classes. These prompts were tailored to guide the recommendation system in different directions, showcasing how even with the same underlying user data, the outputs of the system can vary significantly depending on the design strategy employed. 
\begin{enumerate}
    \item \textbf{Accuracy-Oriented Strategy.} This design approach is focused on delivering high Top$-K$ accuracy. It aims to provide recommendations that align closely with the user's established preferences. To achieve this, we designed \textit{three prompts} in this category, aimed to guide the system to consider a combination of factors from the user profile, including items consumed (watched), favorite genres, and provided user ratings. This approach is tailored to reinforce the user's known tastes and preferences in the recommendations. It could be seen that it can recommend movies \dq{The Incredibles (2004)} (genres: Action|Adventure), \dq{Eternal Sunshine of the Spotless Mind (2004)} -
(genres: Drama|Romance|Sci-Fi), and \dq{the Princess Bride (1987)} - (genres: Adventure|Family|Fantasy), relying on their popularity and relevance to the user's taste. \vspace{0.67mm}

\item \textbf{Beyond-Accuracy-Oriented Strategy.}  The goal of this beyond-accuracy-oriented approach is to broaden the user's viewing experience by introducing \textit{diversity} and \textit{novelty}. It steers away from strictly aligning with known preferences and instead presents a variety of unique and perhaps surprising movie recommendations. It could be noted that for the same given query, this results in the user being recommended a movie like \dq{Moon (2009)} - (genres: Drama|Mystery|Sci-Fi), which may be outside the user's typical viewing history but offers a new cinematic perspective. \vspace{0.67mm}

\item \textbf{Reasoning-Oriented Strategy.} This strategy is centered on enhancing user engagement and understanding. It not only recommends movies but also provides detailed \textit{explanation} and \textit{reasoning} regarding the suggestions. While using these explanations can potentially assist users in exploring and understanding why certain movies are recommended, we are particularly investigating this scenario since previous research has shown that motivating the LLM to reason, for example, through the use of a chain-of-thought (COT) reasoning, could result in more accurate responses from LLMs (i.e., relevant recommendations). An example of recommendation under this strategy could be \dq{The Shawshank Redemption (1994)}, accompanied by an explanation of its themes of hope and redemption that resonate with the motivational depth found in \dq{Toy Story.}
\end{enumerate}

\bluegreen{\subsection{Contributions.} We designed two carefully-structured experiments focusing on classical \textit{top-k recommendation} and \textit{sequential in-context learning}. These experiments involved thousands of API calls to gather the findings presented in this work. The first experiment investigates how different prompt design strategies impact recommendation accuracy, provider fairness, as well as genre dominance and temporal freshness in classical top-k recommendations. The second experiment evaluates the effectiveness and fairness of RecLLMs in sequential in-context learning tasks, comparing zero-shot and few-shot learning scenarios across different datasets. This work makes several contributions to the field of recommender systems, as listed below:
\begin{itemize}
    \item \textbf{Enhanced Analysis of Prompt Design in Zero-Shot RecLLMs:} We provide an in-depth study of how different prompt designs (in terms of intent, structure, and system role) affect the performance of ChatGPT-based recommendation systems. This includes an analysis of various prompt strategies (e.g., accuracy-oriented, beyond-accuracy-oriented, explanation-oriented) and their impacts on recommendation personalization, diversity, recency, and fairness.
    \item \textbf{Identification and Analysis of Biases in Zero-Shot RecLLMs:} Our research identifies and analyzes numerous biases present in RecLLMs, such as item fairness, genre preference bias, and temporal (recency) bias. We compare these biases with traditional CF models, providing insights into how they uniquely manifest in LLM-based systems;
    \item \textbf{Stability Analysis in Zero-Shot RecLLMs:} We study the stability and consistency of recommendations provided by LLM-based systems over time. Specifically, we examine variations caused by changes in underlying data, model updates, and their implications on the quality of recommendations obtained by RecLLMs;
    \item \textbf{Generalization of Fairness Principles in Sequential Recommendations:} We evaluate the effectiveness of RecLLMs in sequential recommendation tasks across different datasets, focusing on zero-shot versus few-shot learning. This analysis provides insights into the adaptability and robustness of these models in maintaining fairness across varied recommendation contexts.
\item \textbf{Impact of User Profile Attributes and Demographic Information on Fairness and Accuracy:} We explore the influence of user profile attributes (historical interaction sampling) and the inclusion of demographic information (e.g., gender, age-group) on recommendation fairness and accuracy, providing new insights into how revealing such information can affect model biases and stereotypes.
\end{itemize}

These contributions collectively enhance our understanding of prompt-based recommendation systems using LLMs, offering new perspectives on their biases, fairness, and stability. Our findings highlight the potential and challenges of integrating LLMs into recommender systems, paving the way for future research and improvements in this domain.}

The structure of the paper is organized as follows. Section \ref{sec:related_work} presents the related work, focusing on the exploration of fairness in recommendation systems (RS) and pre-trained language models (LMs), as well as their applications in enhancing RS. Section \ref{sec:eval} is dedicated to presenting a thorough evaluation of ChatGPT. This is this research's core contribution which we present through the suite of goal-oriented prompts' (Section \ref{subsec:prompts}), the repeated experiments for stability analysis (Section \ref{subsec:rep_exp}), an understanding of the system role in ChatGPT-based RecLLM (Section \ref{subsec:system_role}), and the promotion of fairness (Section \ref{subsec_prom_fair}). Additionally, we explore the explicit versus implicit scenario, which is indicated in prompt structuring (Section \ref{subsec:explicitimplcit}). Following this, Section \ref{sec:exp_setup} outlines the experimental setup employed in our study. The results and key findings of these experiments are thoroughly discussed in Section \ref{sec:results_discussion}. The paper concludes with Section \ref{sec:conc}, where we summarize our findings and outline potential avenues for future research.

\section{Related work}
\label{sec:related_work}

In this section, we briefly review some related work on
recommendation systems and LLM techniques.

\subsection{Fairness in Recommender Systems}

Fairness has become a pivotal topic in AI, gaining scrutiny across branches of trustworthy AI, including security, privacy, and explainability. Fairness in recommender systems has gained significant attention due to their multi-stakeholder nature~\cite{deldjoo2023fairness,ekstrand2019fairness}. Unfairness, even in its minimal form, can adversely impact various stakeholders, including consumers, producers, system designers, supply chains, and even the environment -- the latter are often referred to as~\sq{side-stakeholder}.

The body of literature on fairness in RS is diverse, covering multiple perspectives and dimensions. Recent surveys~\cite{ekstrand2019fairness,deldjoo2023fairness,amigo2023unifying}, categorize these aspects into several dimensions, either orthogonal or partially orthogonal. Key dimensions recognized in the literature include the \textit{main stakeholder} in question (e.g., consumer vs. producer), the target benefits associated with each (such as effectiveness vs. item exposure), the granularity of sensitive groups for assessing fairness (individual vs. group level), and other dimensions including the core definition of fairness, temporal aspects, and more. These dimensions offer a comprehensive view of the fairness landscape in RS, as detailed in these surveys.

To position this study within the literature of fair recommender systems, we introduced Table~\ref{table:research_landscape_multirow}. This table is designed to categorize existing literature along two principal dimensions: the \textit{stakeholder} in question (consumer vs. producer) and the \textit{nature of the core RS} under scrutiny (traditional RS vs. those based on recent advancements in LLMs). We briefly review these dimensions in the following. \vspace{3mm}

\noindent \textbf{Core RS under scrutiny.} We distinguish between \sq{traditional} RS and those enhanced by \sq{RecLLM}. This distinction is crucial because, while RecLLM, such as those based on GPT-like architectures, promise to significantly advance RS landscape with more nuanced and personalized recommendations, they also raise concerns about inherent biases. The extensive and unregulated nature of internet data used to train LLMs raises concerns about biases against specific races, genders, and other sensitive attributes.  For example, if an LLM is trained on e-commerce data where men's products are described more positively than women's products, it might skew recommendations towards male-oriented items. Therefore,~\textit{measuring} these biases is a crucial initial step, and forms the key goal of our current work, in developing effective mitigation strategies. We can briefly review the research on fairness in recommendation systems according to Table \ref{table:research_landscape_multirow} as follows:
\begin{itemize}
    \item \textbf{Traditional RS.} This column lists studies that have focused on traditional methods of recommender systems \cite{deldjoo2018content}. They primarily rely on CF algorithms (possibly using side information of users and items) without the advanced natural language processing capabilities of LLMs. Within these works, some are dedicated to building evaluation frameworks for evaluating RS unfairness, while others focus on developing various mitigation strategies.
\item \textbf{RecLLM}: This column spotlights the burgeoning research domain that merges language models (LMs) with RS representing a significant shift towards harnessing advanced NLP techniques to enhance the accuracy and relevance of recommendations. These studies explore the use of various LMs, including BERT-based models~\cite{shen2023towards} and recent LLMs such as GPT-like architectures~\cite{li2023preliminary,zhang2023chatgpt}. Additionally, beyond the scale of LMs, the target tasks within RS--such as classical recommendation (top-$k$ ranking~\cite{li2023preliminary,zhang2023chatgpt}, sequential), conversational RS~\cite{shen2023towards}, explanation generation, multi-modal recommendations--provide dimensions that could be used for further categorizing these works.
\end{itemize}

\begin{table}[!t]
\caption{Research landscape in Recommender Systems focusing on Consumer and Producer fairness with respect to Age, Gender, Other, and Others}
\centering
\definecolor{Gray}{gray}{0.9}
\begin{tabular}{@{}llcc@{}}
\toprule
 & & \textbf{Traditional RS} & \textbf{RecLLM} \\
\toprule
\multirow{4}{*}{\textbf{Consumer Fairness}} 
& Activity & \cite{li2021user,hao2021pareto,xiao2020enhanced,rahmani2022experiments} & -- \\
& Demographics & \cite{deldjoo2021flexible,deldjoo2021explaining,wu2021fairness,weydemann2019defining,farnadi2018fairness} & \cite{shen2023towards,zhang2023chatgpt,deldjoo2024cfairllm,deldjoo2024fairevalllm} \\
& Merits & \cite{suhr2021does,gomez2021winner}  & -- \\
& Others & \cite{wan2020addressing,lin2021mitigating} & -- \\
\midrule
\addlinespace
\multirow{3}{*}{\textbf{Producer Fairness}}
& Popularity & \cite{dong2021user,da2021exploiting,ge2021towards,zhu2021fairness} & \cite{li2023preliminary} \\
& Demographics & \cite{kirnap2021estimation,boratto2021interplay} &  --\\
& Price/Brand/Location & \cite{deldjoo2021flexible,burke2018balanced,liu2020balancing,shakespeare2020exploring} & -- \\
\midrule
\multirow{1}{*}{\textbf{CP Fairness}}
& Mixed attributes & \cite{naghiaei2022cpfair,rahmani2022unfairness,chakraborty2017fair,patro2020fairrec,wu2021tfrom,do2021two} & -- \\
\bottomrule
\end{tabular}
\label{table:research_landscape_multirow}
\end{table}
\noindent \textbf{Stakeholder.} As mentioned earlier, a major aspect that can be utilized to classify almost all literature on group fairness is the market side focus—whether they concentrate on a single-side market (defined either by consumers or providers) or on both sides \cite{deldjoo2022survey,rahmani2024personalized,naghiaei2022cpfair}. Within each of these segments, as you can observe, we can further categorize the literature based on which sensitive attribute groups are defined. For example, sensitive attributes such as the demographics of consumers (age, gender) and producers, as well as the popularity of items (on the produce side) are quite common focal points.

\begin{itemize}
    \item 
 \textbf{Consumer Fairness.} This category is subdivided into various attributes like Activity, Demographics, Merits, and Others. It includes studies that focus on ensuring fairness among consumers of the recommender system based on these attributes. For instance, ensuring that recommendations are not biased towards a particular demographic consumer group.

 \item \textbf{Producer Fairness.} This focuses on the fairness towards the providers or producers of the content or products recommended by the system. It includes subcategories like Popularity, Demographics, and Price/Brand/Location. These studies might address issues like ensuring lesser-known or niche producers get fair visibility and opportunity in the recommendation process.

 \item 
 \textbf{CP Fairness.} This category involves studies that consider both consumer and producer fairness simultaneously, addressing the balance between the two.
\end{itemize}

It is noteworthy that while traditional Fair-RS research has been extensively explored, RecLLM is an emerging field, presenting its own unique considerations and challenges. Our work is situated in the \sq{Producer Fairness} category under \sq{RecLLM}, focusing on producer fairness in the context of ChatGPT.  It focuses on producer fairness in the context of ChatGPT, particularly examining how \textit{prompt engineering} techniques can be leveraged to address or potentially enhance fairness.

The study by~\citet{zhang2023chatgpt} evaluates the consumer fairness side of zero-shot GPT recommendations, focusing on a variety of consumer demographic attributes but not addressing producer-side fairness. Their work introduces a novel benchmark, FaiRLLM, for evaluating the fairness of RecLLM, highlighting ChatGPT's biases towards certain sensitive user attributes in music and movie recommendations. The work by Deldjoo et al.~\cite{deldjoo2024cfairllm,deldjoo2024fairevalllm}  address the shortcomings by scrutinizing whether changes in recommendations, due to the inclusion of sensitive attributes, result in unfairness. It also explores better ways to normalize and improve the fairness of these models. Conversely, the work by~\citet{li2023preliminary} aligns more closely with ours, focusing on producer unfairness however within the specific domain of \textit{news recommendation}. This study investigates performance of ChatGPT in news recommendation, exploring aspects like personalization, provider fairness, and fake news detection. However, these studies do not address the intricacies of prompt engineering in RecLLMs, nor do they comprehensively address the various forms of biases studies in the current study. Our research goes beyond examining provider fairness based on item unfairness and also considers \textit{other} potential \textit{harms}, such as the recency of recommended items and the stability of recommendations—aspects not explored in the aforementioned studies.

\subsection{Leveraging Pre-trained LMs and Prompting for Recommender Systems}

Recent advancements in recommender systems (RS) have been significantly influenced by the integration of LLMs and innovative prompting strategies.  The use of natural language in recommendation tasks has been explored in various ways. For instance,~\citet{hou2022towards} utilize natural language descriptions and tags as inputs into LLMs to create user representations for more effective recommendations. This contrasts with the narrative-driven recommendations~\cite{bogers2017defining} that rely on verbose descriptions of specific contextual needs.

In terms of prompting techniques, early methods relied on few-shot prompting, where training examples are used as a guide for LLMs~\cite{brown2020language}. With prompt learning, tasks are adapted to LLMs rather than the other way around, utilizing discrete prompts or continuous/soft prompts for task performance. This approach has shown promise across various tasks, including recommendation tasks.

Personalizing LLMs for recommendation is crucial for understanding a user’s intent and addressing their personalized needs. Recent efforts like P5~\cite{geng2022recommendation} and OpenP5~\cite{xu2023openp5} have integrated several recommendation tasks into one LLM using personalized prompts. This approach reformulates recommendation tasks as sequence-to-sequence generation problems, demonstrating the flexibility of LLMs in handling diverse recommendation scenarios. Lastly, prompt transfer research, such as SPoT~\cite{vu2021spot} and ATTEMPT~\cite{asai2022attempt}, focuses on learning from source tasks and utilizing this knowledge for target tasks. This method, including knowledge distillation techniques, shows potential in intra-task prompt distillation and cross-task prompt transfer, contributing to the efficiency and effectiveness of LLM-based recommendation models.

\bluegreen{In recent surveys~\cite{deldjoo2024review,deldjoo2024recommendation}, the wide area of generative models has been extensively explained. These surveys highlight the potential of generative models in transforming traditional RS by leveraging their capabilities in understanding and generating complex data distributions. For instance, they have shown how generative models can effectively handle multimodal data (text, images, and videos), thereby enhancing the richness and accuracy of recommendations. These surveys also shed light on the emergent reasoning abilities of LLMs, such as in-context learning, which allows models to adapt to new tasks with minimal additional training. Moreover, they discuss the application of retrieval-augmented generation (RAG) techniques, which combine information retrieval with generative modeling to produce contextually relevant recommendations. Overall, these insights underscore the significant advancements and ongoing challenges in utilizing generative models for RS.}

\section{Evaluation of ChatGPT-based RecLLM}
\label{sec:eval}
We designed two experiments to evaluate the performance and bias of ChatGPT-based RecLLMs.

\begin{itemize}
    \item \textbf{Experiment 1.} Examining \textbf{prompt design} strategies in classical \textit{top-$k$ recommendation} (cf. Section~\ref{subsec:exp1_designprompt}), and 
    \item \textbf{Experiment 2.} Studying \textbf{generalization} of findings across datasets, and learning strategies (zero vs. few-shot ICL) in \textit{sequential recommendation} task (cf. Section~\ref{subsec:seq_icl1})
\end{itemize}

\noindent Table~\ref{table:experiments_summary} provides a summary of these experiments.

\begin{table}[t!]
    \centering
    \captionsetup{justification=centering, font=small}
    \caption{Summary of Experiments}
    \label{table:experiments_summary}
    \begin{tabular}{@{}p{0.7cm}p{5cm}p{4.5cm}p{4.5cm}@{}}
        \toprule
        \textbf{Exps.} & \textbf{Aim} & \textbf{Approach} & \textbf{Evaluation Metrics} \\
        \midrule
        Exp 1 & (i) Investigate prompt design strategies in classical top-K recommendations (ii) Analyze stability of results over different RecLLM runs, (iii) Look at new content-related metrics (genre diversity, temporal freshness) & Explore prompt formulations (accuracy, diversity, fairness), system role directive, specific information in prompt in the movie domain& Accuracy (top-K precision and recall), beyond-accuracy metrics (diversity, novelty), fairness metrics, genre diversity, temporal freshness \\
        \midrule
        Exp 2 & (i) Assess generalization of fairness across datasets and beyond zero-shot learning to ICL, (ii) Look at the impact of user profile construction and demographic information revealing & Use movie and music dataset, evaluate few-shot and zero-shot ICL in sequential recommendation & Accuracy, provider fairness, catalogue coverage \\
        \bottomrule
    \end{tabular}
\end{table}

\subsection{Experiment 1: Examining Prompt Design Strategies in Classical Top-K Recommendations}\label{subsec:exp1_designprompt}

In the first experiment, we investigate prompt design strategies in a classical top-$K$ recommendation setting. We use an LLM (ChatGPT) as a zero-shot recommender, querying it with various prompt formulations to obtain recommendations. Our goal is to understand how much the recommendation outcome (performance) can be influenced by prompts carrying different objectives (e.g., relevance, increasing diversity, or motivating reasoning). ur goal is to determine if a \textit{one-to-one mapping} exists between the desired natural language query (e.g., enhance diversity) and the actual diversity of the model recommendations.

To achieve the goals of this paper, we examine the relevance and item fairness (including diversity) of the recommendations. Additionally, we explore new aspects of recommendation, such as content diversity and freshness, including genre diversity and temporal freshness.

\begin{table}[h]
\caption{Overview of prompt scenarios designed for the experiment 1.}
\label{tab:prompt_scen}
\centering
\renewcommand{\arraystretch}{2.1}
\begin{tabular}{>{\bfseries}l >{\raggedright\arraybackslash}p{10cm}}
\toprule
 \textbf{Scenario} & \textbf{Description and Prompt} \\
\midrule
\rowcolor[gray]{0.9} \multicolumn{2}{c}{\textbf{Personalization Focused}} \\
\midrule
S1 - Simple & Basic recommendations without additional context. \newline \textit{Prompt: "Recommend 10 movies that the user will likely enjoy."} \\
S2 - Genre-focused & Recommendations focusing on genres and themes similar to the user's past favorites. \newline \textit{Prompt: "Recommend 10 movies that the user will likely enjoy, particularly focusing on genres and themes similar to their past favorites."} \\
S3 - Rating-focused & Incorporation of the user's explicit ratings into the recommendations. \newline \textit{Prompt: "Recommend 10 movies the user will likely enjoy, taking into account both their favorite genres and past movie ratings."} \\
\midrule
\rowcolor[gray]{0.9} \multicolumn{2}{c}{\textbf{Beyond-Accuracy Focused}} \\
\midrule
S4 - Diversify Recommendations & Suggesting lesser-known films to diversify the user's experience. \newline \textit{Prompt: "Suggest 10 high-quality, lesser-known films that diverge from mainstream blockbusters, yet align with the user's tastes."} \\
S5 - Surprise & Offering unexpected recommendations for exploring new preferences. \newline \textit{Prompt: "Offer 10 unique and unexpected movie recommendations aimed at broadening the user’s cinematic horizons beyond their usual preferences."} \\
\midrule
\rowcolor[gray]{0.9} \multicolumn{2}{c}{\textbf{Reasoning-Explanation Focused}} \\
\midrule
S6 - Motivate Reasoning & Providing reasoning for each recommendation to enhance transparency. \newline \textit{Prompt: "Provide 10 carefully selected movie recommendations, each accompanied by a rationale explaining its suitability for the user's preferences."} \\
S7 - Chain of Thought (COT) & Engaging in a logical reasoning process to arrive at the recommendations. \newline \textit{Prompt: "Let's think this through: What would be 10 great movie recommendations for this user and why?"} \\
\bottomrule
\end{tabular}
\end{table}

\subsubsection{Goal-oriented prompts}
\label{subsec:prompts}

Seven distinct prompt scenarios were designed, categorized into three classes focused on personalization, beyond-accuracy metrics, and reasoning. These classes aim to explore how variations in the nature of the prompt influence the recommendations. The scenarios are detailed in Table \ref{tab:prompt_scen} and further exemplified in Section \ref{sec:intro}.

In particular, scenarios \textbf{S1} to \textbf{S3} are focused on core personalization aspects, encompassing basic recommendations without additional context (\textbf{S1}), genre preferences (\textbf{S2}), and the incorporation of explicit user ratings (\textbf{S3}). Meanwhile, Scenarios (\textbf{S4}) and (\textbf{S5}) are designed to enhance the diversity and novelty of the recommendations, with \textbf{S4} emphasizing lesser-known film recommendation and \textbf{S5} aiming to broaden the user's cinematic horizons with unique and unexpected choices. Scenarios \textbf{S6} and \textbf{S7} are explanation-motivation oriented, where \textbf{S6} seeks to provide reasoning for each recommendation to enhance transparency, and \textbf{S7} involves a logical ste-by-step reasoning process to arrive at the recommendations. These scenarios compose a competitive and insightful set of prompts, some of which such as COT, have been successfully tested in other ML disciplines~\cite{wei2022chain,wang2022self,chu2023survey}.\footnote{\textbf{Note.} The CF baselines used in this work operate in an implicit setting. For consistency and fairness, all scenarios in the generative part adhered to this setting, thereby not revealing user movie ratings to the LLM in the prompts. Scenario \sq{Rating-focused} \textbf{S3} is the only exception, included solely for completeness. This is discussed in Section \ref{subsec:explicitimplcit}.}

\subsubsection{Repeated Experiment for the Stability of the Analysis}
\label{subsec:rep_exp}

The experiments were conducted at a temperature setting of 0.0, aiming for predictable responses. However, we observed variations in the recommendations, possibly due to factors like trending data, recent updates in the database, or changes in the behavior of ChatGPT. This suggests that the recommendation system may interpret the same input differently in each iteration, independent of the temperature setting.

To assess the robustness and reliability of our system, we repeatedly conducted recommendation queries for each user, five times each, to examine if outputs vary despite consistent input.  We further adopted a \textit{bootstrapping sampling strategy} during the evaluation stage, as detailed in Section \ref{subsec:bs}. This method is particularly effective in revealing the reliability of the results, even when dealing with datasets that are small or may not perfectly represent the entire population. This thorough evaluation approach, therefore, enhances our confidence in the performance of the system across various potential real-world scenarios.

\subsubsection{Understanding the impact of \dq{System} Role in ChatGPT}
\label{subsec:system_role}

In this study, we sought to understand the impact of different system roles assigned to ChatGPT on recommendation outcomes. We focused on assessing whether responses significantly differ when specific \dq{system} instructions are attached, and how these variations influence the functionality of a recommender system. Specifically, we aimed to determine the effectiveness of embedding fairness directly into the system role versus incorporating it in the prompt itself. Table \ref{tab:roles} summarizes the roles assigned to ChatGPT during our experiments:

\begin{table}[!h]
\caption{Summary of Different Roles Assigned to ChatGPT during experiments\label{tab:roles}}
\centering
\renewcommand{\arraystretch}{1.5}
\begin{tabular}{>{\bfseries}l >{\raggedright\arraybackslash}p{10cm}}
\toprule
 \textbf{Role ID} & \textbf{Description} \\
\midrule
R0 - No Role & Direct user-driven prompts without system context. \\
R1 - System Role as Recommender & System-centric role focusing purely on user information. \\
R2 - System Role as Fair Recommender & System role with an explicit fairness objective. \\
\bottomrule
\end{tabular}
\end{table}
\subsubsection{Fairness Emphasis}
\label{subsec_prom_fair}

We also aim to understand whether integrating a fairness statement in the system or in the prompt itself is more effective. To assess the impact of explicit directives on the recommendation system, each scenario was optionally combined with a \sq{Fairness Emphasis Statement.} These statements directed the model towards specific objectives, here ensuring fairness. The considered emphasis options included

\begin{itemize}
\item \textbf{E0 - Without Fairness Statement:} No fairness statement is included in this option. Prompts are used as described in Table \ref{tab:prompt_scen}.
\item \textbf{E1 - With Fairness Statement:} This option emphasizes ensuring fair recommendation between popuar nd less popular movies. \dq{\textit{Prompt: Ensure a fair representation of both popular and less-known movies. Based on these movies: \{user\_movies\_string\}, recommend 10 movies that the user will likely enjoy}}. The latter prompt is combining Scenario (S1/E1).

\end{itemize}


\label{subsec:exp2_fewshot}
\subsubsection{Explicit vs. Implicit Scenario}
\label{subsec:explicitimplcit}
This test evaluated whether displaying the movies a user has rated, as indicated in \{user\_movies\_string\}, enhances the quality of RS, based on the premise that including such information could improve recommendations. Scenarios \textbf{S1 to S6}, except \textbf{S3}, were conducted implicitly, without user ratings in the prompts, mirroring the implicit mode of the CF baseline experiments. Uniquely, \textbf{S3} was examined in both explicit and implicit contexts. In the explicit context, user ratings were integrated into the prompts, but were absent in the implicit context. This explicit approach was specific to \textbf{S3}. An example of \{user\_movies\_string\} and the presentation of user movie lists and ratings in the explicit context is as follows. \vspace{1.5mm}

\noindent \textbf{Example.} To illustrate the experimental scenarios, consider a user's movie list represented in this manner.

\begin{center}
\begin{minipage}{0.8\textwidth}
\begin{verbatim}
The Matrix (Genres: Action|Sci-Fi), Inception (Genres: Action|Thriller)
\end{verbatim}
\end{minipage}
\end{center}

\noindent In the Explicit scenario, the prompt would be constructed to include these ratings, exemplified here:

\begin{center}
\begin{minipage}{0.8\textwidth}
\begin{verbatim}
The Matrix (Genres: Action/Sci-Fi, Rating: 5/5), Inception (Genres: Action/Thriller, Rating: 4.5/5)
\end{verbatim}
\end{minipage}
\end{center}

\vspace{1.8mm}

The primary hypothesis under investigation is whether the inclusion of explicit ratings enables the system to tailor recommendations more closely to the user's demonstrated preferences, thereby enhancing the personalization of the recommendation process.

\begin{figure}[!t]
    \centering
    \includegraphics[width =0.98\linewidth]{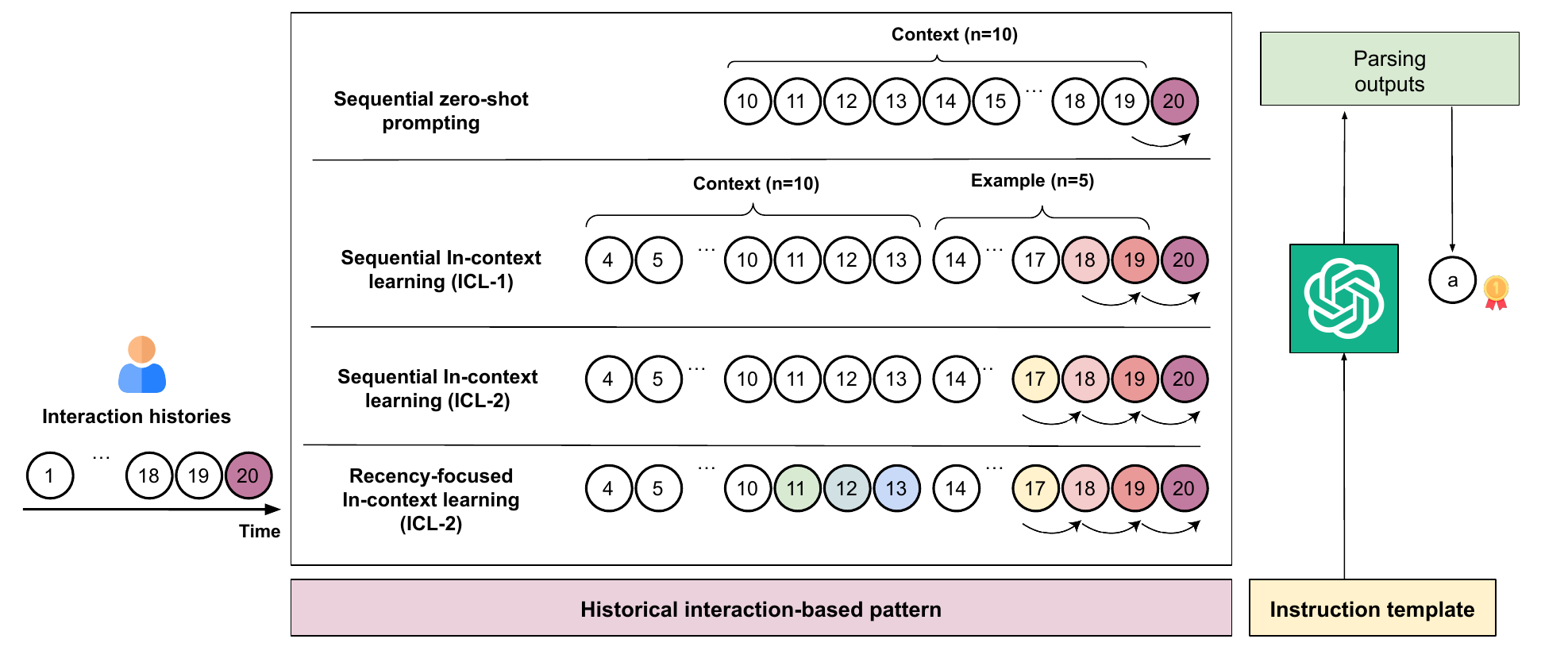}
\caption{Sequential in-context learning for various scenarios explored in Experiment 2 of this research.}
    \label{fig:ICL}
\end{figure}
\bluegreen{\subsection{Experiment 2: Sequential In-Context Learning}
~\label{subsec:seq_icl1}
The task of recommending suitable movies/music to users can be modeled as a sequential ranking problem where the historical interactions of a user are utilized to predict future preferences. In this context, the goal is to rank candidate items \( C = \{i_j\}_{j=1}^{m} \) such that the most relevant items for the user appear at the top. Given a user's interaction history \( H = \{i_1, i_2, \ldots, i_n\} \), to leverage it for generating recommendations, we opt to divide it into two main parts, as displayed in Figure \ref{fig:ICL}:

\begin{enumerate}
    \item \textbf{Context part.} This part is sampled from the~\textit{entire interaction history} of the user to provide general context about the user's interests, passion, and preference (toward music and movies). We use two strategies for sampling:
    \begin{itemize}
        \item \textit{Frequent/Highest-rated:} Selects the most enjoyed items by the user, based on the ratings provided (for movies) or frequency of consumption (for music);
        \item \textit{Recency-Focused:} Utilizes a weighted average method to emphasize recently consumed items. This strategy selects items that have not only been liked by the user but also have been recently listened to or watched, ensuring relevance and timeliness in the recommendations.
    \end{itemize}
    \item \textbf{Example part.} This part is taken from~\textit{recent interactions} to guide specific next recommendations. It directly determines the next items to be consumed in sequential tasks and is used in ICL-$1$, and ICL-$2$, which denote the \textit{few-shot learning} scenarios, using one or two examples.
\end{enumerate}

\noindent Based on this division, we explore different prompting strategies, as outlined in the following.  We note that Experiment 2 has been inspired by~\citet{hou2024large} and extends it by exploring the above profile investigation, and evaluation going beyond pure accuracy. \vspace{2mm}

\noindent \textbf{Note.} In the prompting scenarios outlined below, we \textit{optionally} incorporate \dq{user demographic information} (such as gender, or age category) into the prompts to examine how revealing this information might influence model biases and stereotypes. Specifically, where we have \texttt{\{is demographic\}} information, it is used in statements like \texttt{\dq{The user is female}}, \texttt{\dq{The user is young}}. Details of the results of comparisons with and without demographic information on fairness are provided in Table \ref{tab:sim_align}.}


\bluegreen{\subsection*{Sequential Zero-Shot Prompting}
\label{subsec:seq_zs}
This method only uses the context part of the user interaction history}

\begin{quote}
\textbf{Example:} 
\begin{verbatim}
The user {is female, and} has watched the following movies:

- "From Russia with Love (1963)" in genre(s) Action with rating 5
- "Star Trek IV: The Voyage Home (1986)" in genre(s) Action|Adventure|Sci-Fi with rating 4
- "Planet of the Apes (1968)" in genre(s) Action|Sci-Fi with rating 3
- "Star Wars: Episode I - The Phantom Menace (1999)" in genre(s) Action|Adventure|Fantasy|Sci-Fi with rating 4
- "Final Conflict, The (a.k.a. Omen III: The Final Conflict) (1981)" in genre(s) Horror with rating 3

This selection reflects the user's movie preferences.
What would be the top-1 suitable next recommendation?
\end{verbatim}
\end{quote}

\noindent \bluegreen{as stated earlier, \texttt{\{is female\}} is essentially denoted with \texttt{\{is demographic\}} below, which is tested in two scenarios: \textit{included} and \textit{omitted}.}

\bluegreen{\subsection*{Sequential In-Context Learning (ICL-1)}

This approach takes the contest part of the preivous part, but taked the xample part of the interaction and from that take the last interatcion to guide the model how the recommendation. The examples are derived from recent interactions.}
\begin{quote}
\textbf{Example:} 
\begin{verbatim}
The user {is demographic, and} has watched the following movies:

- "From Russia with Love (1963)" in genre(s) Action with rating 5
- "Star Trek IV: The Voyage Home (1986)" in genre(s) Action|Adventure|Sci-Fi with rating 3
- "Planet of the Apes (1968)" in genre(s) Action|Sci-Fi with rating 3
- "Star Wars: Episode I - The Phantom Menace (1999)" in genre(s) Action|Adventure|Fantasy|Sci-Fi with rating 4
- "Final Conflict, The (a.k.a. Omen III: The Final Conflict) (1981)" in genre(s) Horror with rating 3

This selection reflects the user's movie preferences.
Given the user has recently watched the following movies in order:
1. "Face/Off (1997)" in genre(s) Action|Sci-Fi|Thriller with rating 4
2. "Bringing Out the Dead (1999)" in genre(s) Drama|Horror with rating 5
3. "Sixth Sense, The (1999)" in genre(s) Thriller with rating 5
4. "Austin Powers: The Spy Who Shagged Me (1999)" in genre(s) Comedy with rating 4
5. "Arlington Road (1999)" in genre(s) Thriller with rating 4

You should recommend:
Recommendation 1: "Bowfinger (1999)" in genre(s) Comedy with rating 3
What would be the top-1 suitable next recommendation after the above movies?

\end{verbatim}
\end{quote}

\bluegreen{\subsection*{Sequential In-Context Learning (ICL-2)}
\label{subsec:seq_icl2}

Similar to ICL-1 but with a bigger selection of examples equal to 2. The examples are chosen to provide more specific guidance on generating recommendations based on recent user interactions.}
\begin{quote}
\textbf{Example:} 
\begin{verbatim}
The user {is demographic, and} has watched the following movies:

- "From Russia with Love (1963)" in genre(s) Action with rating 5
- "Star Trek IV: The Voyage Home (1986)" in genre(s) Action|Adventure|Sci-Fi with rating 3
- "Planet of the Apes (1968)" in genre(s) Action|Sci-Fi with rating 3
- "Star Wars: Episode I - The Phantom Menace (1999)" in genre(s) Action|Adventure|Fantasy|Sci-Fi with rating 4
- "Final Conflict, The (a.k.a. Omen III: The Final Conflict) (1981)" in genre(s) Horror with rating 3

This selection reflects the user's movie preferences.

Given the user has recently watched the following movies in order:
1. "Strange Days (1995)" in genre(s) Action|Crime|Sci-Fi with rating 3
2. "Face/Off (1997)" in genre(s) Action|Sci-Fi|Thriller with rating 4
3. "Bringing Out the Dead (1999)" in genre(s) Drama|Horror with rating 5
4. "Sixth Sense, The (1999)" in genre(s) Thriller with rating 5
5. "Austin Powers: The Spy Who Shagged Me (1999)" in genre(s) Comedy with rating 4

You should recommend:
Recommendation 1: "Bowfinger (1999)" in genre(s) Comedy with rating 3
Recommendation 2: "Arlington Road (1999)" in genre(s) Thriller with rating 4
What would be the top-1 suitable next recommendation after the above movies?
\end{verbatim}
\end{quote}

\subsection*{Recency-Focused Zero-shot and ICL-1 and ICL-2}
\label{subsec:rec_foc}
\bluegreen{We replicate all previous prompting scenarios but modify the \dq{context part} to include the most recent interactions. This approach indicates to the ChatGPT recommender system that the best recommendations are based on the user's recent consumption.}

\vspace{4mm}

\noindent The next sections outlines experiments conducted to assess various reproducibility aspects, including the obtained results and our observations.

\section{Experimental Setup}
\label{sec:exp_setup}
\bluegreen{In this section, we detail the experimental setup for two experiments, including the \textit{datasets,} \textit{evaluation metrics,} \textit{bootstrapping methods,} \textit{baselines employed,} and \textit{hyperparameter tuning}.}
\subsection{Experiment 1. Classical Top-$k$ Recommendation}
\label{subsec:exp1_topk}
\subsubsection{Datasets and Setup}
\bluegreen{Experiment 1 used the \texttt{MovieLens-Latest-Small} dataset from the GroupLens group.\footnote{\url{https://grouplens.org/datasets/movielens/}} This dataset comprises the latest movie entries and was used in its entirety to carry out experiments in part 1 of our study. The experimental design required conducting API calls for each user, resulting in a total of 610 API calls (equal to the number of users) per individual recommendation setting. Given the scope—610 users, 7 different prompts or prompt scenarios (cf. Section~\ref{subsec:prompts}), 2 fairness statements in the prompt (cf. Section~\ref{subsec_prom_fair}), and 2 system roles (cf. Section~\ref{subsec:system_role})—we generated \textbf{thousands} of scenarios or API calls. Due to the computational demands, we limited our experiments in part 1 to a \textit{single dataset}, focused on the \textbf{classical top-$k$ recommendation task.} The detailed statistics of the dataset are provided in Table~\ref{tbl:datasets}, where we implemented a data split of 80\% training, 10\% validation, and 10\% testing. The validation dataset is used solely for CF baseline hyperparameter tuning, while both RecLLMs and the CF baseline utilize the same training and testing datasets to conduct recommendations.}

\subsubsection{Evaluation Metrics}
\label{subsec:eval_metrics}
\bluegreen{The first experiment employs a diverse range of metrics to assess the performance of different RS. These metrics are categorized into two primary classes: \textit{accuracy metrics}, and \textit{provider fairness/diversity}. Additionally, we incorporate new \textbf{content-centric metrics} such as \textit{temporal freshness} and \textit{genre dominance/tendancy} to explore insights into potential biases and tendencies of RecLMMs compared to mainstream CF models.} 

\subsubsection*{Accuracy Metrics.} These conventional metrics measure the quality of the top-$k$ recommendation or the personalization of recommendations 
\begin{itemize}
    \item \textbf{NDCG@k:} Measures the ranking quality by evaluating the placement of relevant items within the top-$k$ positions.
    \item \textbf{Recall@k.} The proportion of relevant items that are successfully recommended.
\end{itemize}
High scores in these metrics indicate more accurate and relevant recommendations. \vspace{2mm}

\noindent \textbf{Provider Fairness/Diversity.}
\bluegreen{Beyond conventional accuracy metrics, these metrics evaluate the fairness of the recommendations from the provider perspective. They assess how well the system introduces new and varied content and how equitably it treats different item providers.

\begin{itemize}
    \item \textbf{Gini Index~\cite{deldjoo2021explaining}:} Measures the inequality in the distribution of item recommendations. A lower Gini Index indicates a more equitable distribution among items. The Gini index is calculated as follows:
    \[
    \text{Gini} = \frac{\sum_{i=1}^{n} (2i - n - 1) x_i}{n \sum_{i=1}^{n} x_i}
    \]
    where \( x_i \) represents the number of times item \( i \) is recommended, sorted in ascending order, and \( n \) is the total number of items. The Gini index becomes 0 when all items are recommended an equal number of times, indicating perfect equality. Mathematically, if all \( x_i \) are equal, then:
    \[
    x_i = \frac{\sum_{i=1}^{n} x_i}{n} \quad \forall i \implies \text{Gini} = 0
    \]
    The Gini index becomes 1 when one item is recommended exclusively, indicating maximum inequality. In our work, we interpret the Gini index to measure item fairness, \textit{where lower Gini values indicate higher (better) item fairness.}
    
    \item \textbf{HHI (Herfindahl-Hirschman Index)~\cite{HHI2018,wang2022make}:} This metric 
    Assesses the concentration of recommendations among items. It is calculated as follows:
    \[
    \text{HHI} = \sum_{i=1}^{n} \left( \frac{x_i}{\sum_{j=1}^{n} x_j} \right)^2
    \]
    The HHI value ranges from $1/n$ to 1, where \( n \) is the total number of items. The HHI is $1/n$ when all items are recommended equally, indicating maximum fairness, and it is 1 when one item is recommended exclusively, indicating no fairness. Here, \textit{lower values of the HHI are considered indicative of better item fairness}.
    
    \item \textbf{Entropy:} Reflects the diversity and unpredictability of recommendations. Higher entropy values indicate greater diversity. The Entropy is calculated as follows:
    \[
    \text{Entropy} = -\sum_{i=1}^{n} p_i \log(p_i)
    \]
    where \( p_i = \frac{x_i}{\sum_{j=1}^{n} x_j} \) and \( p_i > 0 \). The entropy value ranges from 0 to \(\log(n)\), where 0 indicates no diversity (one item recommended exclusively) and \(\log(n)\) indicates maximum diversity (all items recommended equally). \textit{Higher values of Entropy are considered indicative of better item fairness}.
\end{itemize}

\noindent While the above metrics are used to measure item fairness based on item exposure, they also serve as indicators of how recommendations produce \textbf{diverse} outcomes.}

\subsubsection{Bootstrapping Sampling Strategy}
\label{subsec:bs}

In experiment 1, for the evaluation, we additionally employ a~\textit{bootstrapping sampling strategy,} a statistical technique where 1000 samples were generated by repeatedly resampling our dataset with replacement \cite{deldjoo2023fairnessGPT}. This method was chosen to enhance the robustness and reliability of our analysis, particularly in measuring the performance metrics of our recommendation system, such as NDCG, Recall, and Precision.

Through this approach, we computed means for each metric across all bootstrap samples, then calculated their averages and 95\% confidence intervals. This strategy is implemented for understanding the variability and confidence in our metrics, to ensure our evaluation reflects the likeness of the system performance on real-world performance. This is especially beneficial in situations with small or non-representative samples, allowing for more reliable population inferences.
\subsubsection{CF Baselines.}
The choice of models and the corresponding hyperparameter optimization process is outlined as follows.

\begin{table}
\caption{Statistics of the final datasets used in this work after $k$-core pre-processing.}
\label{tbl:datasets}
\centering
\begin{tabular}{lcccccccc}
\toprule
\textbf{Dataset} & |U| & |I| & |R| & $\frac{R}{U}$ & $\frac{R}{I}$ & Density & Item Gini & User Gini \\
\midrule
\textbf{MovieLens Latest Small}  & 610 & 9,724 & 100,836 & 165.30 & 10.36 & 98.3\% & 0.715 &0.603 \\
\bottomrule
\end{tabular}
\end{table}

\subsubsection*{Models.}
We use a suite of competitive, CF recommendation models as baseline ranking models in our post-processing approach, as summarized below. 

\begin{itemize}
    \item \textbf{BPR} \cite{rendle2012bpr}: A conventional recommendation model that employs matrix factorization to learn user and item embeddings of low dimensionality, and optimize the model based on the pairwise ranking of items for each user to predict whether a user prefers a given item over another;
        \item \textbf{ItemKNN} \cite{sarwar2000analysis}: An item-based K-nearest employs neighbors algorithm, utilizing similarity metrics such as cosine similarity to identify the closest item neighbors. These neighboring items are then leveraged to predict scores for user-item pairs.

    \item \textbf{MultiVAE} \cite{liang2018variational}: A non-linear probabilistic deep learning model that extends a variational autoencoder (VAE) structure to collaborative filtering for implicit feedback, and acquires the underlying representations of users and items from their interactions to create recommendations in an unsupervised way.

    \item \textbf{LightGCN} \cite{he2020lightgcn}: A pure collaborative filtering method that utilizes a simplified version of graph convolutional networks (GCNs) without nonlinear activation functions and additional weight matrices. It learns user and item embeddings through graph propagation rules and user-item interactions, making it scalable and efficient.

    \item \textbf{NGCF} \cite{wang2019neural}: A graph-based recommendation model that employs a neural network architecture and learns high-order connectivity and user-item signals based on the exploitation of the user-item graph structure, by propagating embeddings on it. 

\end{itemize}

The selected CF models constitute a set of competitive baselines from various natures, representing a diverse array of recommendation approaches (classical, neural, graph-based). Additionally, the \textbf{TopPop} method is included as a non-informative baseline, as an essential baseline to provide a case where the outcomes are heavily biased towards popular items.

\subsubsection*{Hyperparameter Tuning.}

In our research, the primary focus is on the hyperparameter tuning of various collaborative filtering (CF) recommendation models, selected as baselines. These models are specifically tailored and evaluated within implicit feedback scenarios, where the subtleties of user preferences are inferred from indirect interactions. This approach ensures a comprehensive exploration of hyperparameter spaces to optimize model performance in these nuanced environments. The RecBole public library\footnote{\url{https://www.recbole.io/}} is used for implementing and applying these models. Hyperparameter tuning is an essential step in optimizing the performance of these models. We used a bootstrapping sampling strategy, generating 1000 samples by repeatedly resampling our dataset with replacement.

The following is a summary of the hyperparameter tuning conducted for each model:
\vspace{2mm}
\begin{itemize}
    \item \textbf{BPR-MF.} \verb|Embedding size| choices \verb|'[32, 64, 128]'| and \verb|learning rate| choices \verb|'[1e-4, 5e-4, 1e-3, 5e-3]'|. Total of 12 cases.
    \item \textbf{ItemKNN.} \verb|k| choices \verb|'[10, 50, 100, 200, 250, 300, 400]'| and \verb|shrink| choices \verb|'[0.0, 0.1, 0.5, 1, 2]'|. Total of 35 cases.
    \item \textbf{NGCF.} \verb|Learning rate| choices \verb|'[1e-4, 5e-4, 1e-3]'|, \verb|hidden size list| choices \verb|'["[64, 64, 64]", "[128, 128, 128]"]'|, \verb|node dropout| choices \verb|'[0.0, 0.1, 0.2]'|, \verb|message dropout| choices \verb|'[0.0, 0.1, 0.2]'|, and \verb|regularization weight| choices \verb|'[1e-5, 1e-3]'|. Total of 108 cases.
    \item \textbf{MultiVAE.} \verb|Learning rate| choices \verb|'[1e-4, 5e-4, 1e-3, 5e-3]'| and \verb|latent dimension| choices \verb|'[64, 128, 200, 300, 512]'|. Total of 20 cases.
    \item \textbf{LightGCN.} \verb|Learning rate| choices \verb|'[5e-4, 1e-3, 2e-3]'|, \verb|number of layers| choices \verb|'[1, 2, 3, 4]'|, and \verb|regularization weight| choices \verb|'[1e-05, 1e-04, 1e-03, 1e-02]'|. Total of 48 cases.
\end{itemize}

The total number of hyper-parameters is 14, encompassing a total of 223 experimental cases. This approach ensures that we effectively explore a wide range of hyperparameter configurations, particularly for models operating in implicit feedback scenarios, which are crucial for our research.

\subsection{Experiment 2. Sequential Recommendation}
\subsubsection{Datasets and Setup}
\bluegreen{In this study, our objective is to understand the~\textit{generalizability} of the findings from the previous study and address aspects that were not studied previously. In particular, Experiment 2 focuses on the more practical task of \textbf{sequential recommendation} and the comparison of~\textit{zero-shot} vs.~\textit{few-shot in-context learning} (ICL), which we evaluate on datasets of different domains (movies and music). The choice of sequential recommendation, rather than classical \textbf{top-$k$}, was motivated by the research carried out in \cite{hou2023large}. Therefore, the primary goal in Experiment 2 is not to assess the quality of different prompt scenarios (as carried out in Section~\ref{subsec:exp1_topk}), rather~\textit{to evaluate the effectiveness of zero-shot vs. few-shot ICL  on the accuracy vs. item fairness of RS.}}

\bluegreen{We focus our attention on \underline{two} main datasets from the movie and music domains:~\textit{MovieLens-1M} and~\textit{LastFM-1K.} These datasets are widely recognized in the academic community, and provide a basis for evaluating our models in terms of fairness across different types of media consumption data. Here, we opt to use a sub-sample of users instead of all, to speed up experiments. Initially, we randomly select a subset of~\textit{80 users} who exhibit a moderate level of interaction within the datasets. This allows us to handle the data efficiently while ensuring that the users selected have enough interactions to inform the training process. The data for these users is divided into training and test sets by sorting their interactions over time and splitting them so that 80\% of a user interactions are used for training, with the remaining 20\% held out for testing. This method respects the~\textit{chronological order} of interactions, thereby simulating a realistic scenario where a model can only learn from past data to make predictions about future user behavior.}

\begin{table}[!t]
\caption{Statistics of the final datasets used in the sequential ICL task.}
\label{tbl:datasets_2}
\centering
\begin{tabular}{lcccccccc}
\toprule
\textbf{Dataset} & |U| & |I| & |R| & $\frac{R}{U}$ & $\frac{R}{I}$ & Density & Item Gini & User Gini \\
\midrule
\textbf{LastFM-1K}  & 80 & 5,500 & 277,607 & 3,470.09 & 50.47 & 98.6\% & 0.697 & 0.621 \\
\textbf{MovieLens 1M}  & 80 & 3,900 & 20,987 & 262.34 & 5.38 & 99.2\% & 0.745 & 0.657 \\
\bottomrule
\end{tabular}
\end{table}

\bluegreen{\subsubsection{Baselines.} In this experiment, we do not compare the built RecLLMs with traditional sequential models, as this would require a separate and labor-intensive set of experiments, which is beyond the scope of this study. Instead, we compare the fairness and accuracy of different versions of RecLLMs against each other. Additionally, in one part of the experiment, we explore the impact of the presence or absence of \textit{user demographics} in the prompts on recommendation fairness and quality. Since CF baselines typically do not use such demographic information, we exclude them from this comparison.}

\subsubsection{Evaluation metrics}

\bluegreen{The first experiment employs a diverse range of metrics to assess the performance of our recommendation system. These metrics are categorized into two primary classes: \textit{accuracy metrics}, and \textit{item fairness}, and \textit{catalogue coverage}. 

\subsubsection*{Accuracy Metrics.} For accuracy, we choose \underline{simpler} metrics since the goal is not to compare them with strong CF baselines. We use the following two metrics to measure the quality of the top-$k$ recommendations in sequential setting:
\begin{itemize}
    \item \textbf{Hit Rate (HR@k):} Measures the proportion of users for whom at least one of the top-$k$ recommended items is relevant. It is calculated as:
    \[
    \text{HitRate@K} = \frac{|\{i \in R_Q : i \in \mathcal{G}\}|}{K}
    \]
    where \( R_Q \) represents the set of recommendations of the ranker in question and \( \mathcal{G} \) represents the set of items in the ground truth.
    
    \item \textbf{Average Rank (AverageRank@K):} Measures the average position of relevant items in the recommendation list. It is calculated as:
    \[
    \text{AverageRank@K} = \frac{1}{|\{i \in R_Q : i \in \mathcal{G}\}|} \sum_{i \in \mathcal{G}} \text{rank}_{R_Q}(i)
    \]
    where \( \text{rank}_{R_Q}(i) \) is the rank of item \( i \) in the recommendation list \( R_Q \).
\end{itemize}
High scores in these metrics indicate more accurate and relevant recommendations.} \vspace{2mm}

\subsubsection*{Item fairness, catalouge Coverage and long-tail distribution.}
\bluegreen{Similar to the previous experiment, we use the \textbf{Gini Index}, \textbf{HHI}, and \textbf{Entropy} to measure item fairness. Additionally, we compute \textbf{catalogue coverage}, and display the \textbf{long-tail distribution} to visualize the concentration bias of recommendated items.}

\bluegreen{\subsection{Parsing Output Recommendations and Finding Closest Match Items}

Since language models (LLMs) naturally produce text-based outputs, we need a systematic approach to parse these outputs and map them to our existing dataset items. This ensures that when the RecLLM generates recommendations for which we do not find \textit{an exact match} in the dataset, we can associate them with the most similar existing items, maintaining the integrity and relevance of our recommendation system.

\begin{itemize}
    \item We start by utilizing the OpenAI GPT-3.5 API itself to parse the raw text of the recommendations. We designed a \texttt{LLM\_parsing} function to extract the recommended songs and artists (or movies) from the text provided by the language model. For instance, we specify a system message to format the output in a structured way, ensuring that each recommendation follows the format: ``\textit{Song Name} by \textit{Artist}.'' This structured output helps the subsequent matching process. The parsed recommendations are then collected, which contain cleaned and structured text.

\item Once the recommendations are parsed, the next step involves finding the closest matches to these recommendations within our dataset. This is achieved using the \texttt{find\_closest\_match} function, which employs the \texttt{difflib.SequenceMatcher} from the Python standard library. This tool calculates \textbf{similarity scores} between the recommended song and artist names and the existing items in our dataset. For each recommendation, the function compares it against all items in the dataset, computing a combined score based on the similarity of both the track name and the artist name.

\item A threshold is set to ensure that only matches with a high enough similarity score are considered valid. If the highest score exceeds this threshold, the corresponding item ID from the dataset is selected as the closest match. This method ensures robustness by preventing low-similarity matches from being accepted, thus maintaining the relevance and accuracy of the recommendations.
\end{itemize}

In summary, the implementation of these methods leverages essential Python tools and libraries, \texttt{openai} for API interaction, and \texttt{difflib} for sequence matching. These tools provide a robust and efficient framework for parsing and matching recommendations, ensuring that the RecLLM system can be effectively evaluated even when it generates items not originally present in the dataset. 
}

\section{Results and Discussion}
\label{sec:results_discussion}

In this section, we present the results and discussion for two experiments. We formulate hypotheses for each part, then showcase the results and provide analysis. Finally, we answer the hypotheses. 


\subsection{Experiment 1. Zero-Shot Top-k Recommendation} Through experiment 1, conducted in a zero-shot setting for the classical top-$k$ recommendation, we aim to answer the following experimental research questions.  Details about Experiment 1 can be found in Section \ref{subsec:exp1_designprompt} and Section \ref{subsec:exp1_topk}.

We evaluate the average performance (Avg Perf.) in terms of NDCG and Recall across different conditions/models. We compare regular performance values (average) with bootstrap averages to gain deeper insights. The bootstrap method involves repeated resampling and offers a more robust understanding of performance variability. Table~\ref{tbl:results} illustrates this comparison. While the regular average provides a single estimate, the bootstrap mean extends this by offering confidence intervals, indicating the range within which the true performance measure is likely to fall.

\begin{tcolorbox}[colback=white, title= Experimental Research Questions , colframe=black, colbacktitle=gray!85!black]

\textbf{RQ1.} How does the incorporation of various goal-oriented prompts impact the \textbf{accuracy} (personalization) of GPT-based model in comparison to CF baselines? \textit{(cf. Section \ref{subsec:personal})}

\vspace{2mm}

\textbf{RQ2.} What is the \textbf{stability} and \textbf{consistency} of the personalization performance metrics (e.g., NDCG, Recall) for GPT-based recommendation systems across multiple runs? How does this variability compare to CF baselines? \textit{(cf. Section \ref{subsec:stable})}

\vspace{2mm}

\textbf{RQ3.} How do GPT-based recommendation systems compare to CF baselines in terms of provider fairness? Can the inclusion of a \dq{Fair Recommender} system role might mitigate item unfairness and enhance diversity? \textit{(cf. Sections \ref{subsec:prov_fairness}})

\vspace{2mm}

\textbf{RQ4.} Do GPT-based recommendation systems exhibit a bias towards recommending newer or older movies, or certain genre tendancy, compared to CF baselines, and how does the temporal freshness of recommendations hold? How does the inclusion of explicit user ratings in prompts affect the personalization performance? \textit{(cf. Sections \ref{subsec_temporalFesh})}.

\vspace{2mm}

\textbf{RQ5.} How do different algorithmic in-context learning (ICL) strategies, --- in particular zero-shot vs. few-shot learning -- impact the quality and biases of RecLLMs? What are the specific prompt design aspects that may influence this towards better or worse performance? \textit{(cf. Section \ref{subsec:seq_rec})}

\vspace{2mm}

\textbf{RQ6.}  What are the economic and practical implications of using GPT-based models for recommendation systems, specifically in terms of inference costs and latency issues? \textit{(cf. Section \ref{subsec:scaliabity})}

\end{tcolorbox}

\subsubsection{Personalization of Recommendations}
\label{subsec:personal}

This section focuses on the personalization performance measured by recommendation top-$K$ accuracy. Results are presented in Table~\ref{tbl:results}, with each generative scenario (using ChatGPT) in the upper sections and CF baselines in the bottom section. Not every possible scenario combination was tested; rather, a select subset was examined.

\subsubsection*{Hypotheses} To investigate the effectiveness of different recommendation strategies, we propose the following hypotheses:

\begin{itemize}
    \item \textbf{H1.} Incorporating goal-oriented prompts, including \textit{personalization-based}, \textit{beyond-accuracy}, and \textit{reasoning-based prompts} (as detailed in scenarios \textbf{S1 to S7} in Section \ref{subsec:prompts}), with different system assignment roles, results in substantially varied performance outcomes in RecLLMs; \vspace{2mm}

\item \textbf{H2.} In terms of performance, GPT-based recommenders in~\textit{zero-shot setting} can provide comparable performance compared to collaborative filtering (CF) baselines.
\end{itemize}

\subsubsection*{Discussion.}

Our evaluation highlights several findings regarding the personalization of recommendations using RecLLM. 

\begin{itemize}
    \item First, there is a clear indication that incorporating the \dq{system role} improves the performance of RecLLM, and this consistency is observed across most scenarios. For example, for a randomly chosen scenario S1, in \textbf{Tab1} and \textbf{Tab2} (with no system role assignment), the NDCG performances are 0.008803 and 0.010431, respectively. In contrast, \textbf{Tab3} and \textbf{Tab4} (with system role assignments, either as \dq{act as a recommender system} or \dq{act as a fair recommender system}) show performance improvements to 0.013007 and 0.014268, respectively. This represents roughly a 45\% improvement. In some scenarios, the improvement reaches up to 100\%. It becomes thus evident that \textit{the system role plays a crucial part in optimizing the effectiveness of recommendation systems}. \vspace{1mm}

   \item Second, the performance varied considerably across different scenarios (S1 to S7), indicating that the context and nature of the prompts affect recommendation outcomes. For example, in \textbf{Tab4}, the \dq{Simple} prompt scenario (S1) achieved an NDCG of 0.014, while the \dq{Diversify} prompt (S4) had a lower NDCG of 0.002. The best-performing methods were \textbf{simple}, \textbf{reasoning-based} (COT and motivated reasoning), providing relatively better accuracy. Thus, based on these results, we may conclude that \textit{requesting ChatGPT to diversify recommendation or to provide novel recommendations significantly reduces accuracy, which should be avoided or carefully considered.} Overall these variations underscore the importance of selecting appropriate prompts based on the desired recommendation attributes. \vspace{1mm}
\item Finally, comparing RecLLMs with baselines, our results reveal that RecLLM, even in zero-shot settings, generally performs lower than CF baselines, though in certain contexts, this difference shrinks considerably. For instance, the average NDCG of CF models such as BPR-MF and ItemKNN were 0.04776 and 0.04905, respectively. In comparison, the best-performing RecLLM prompt achieved an NDCG of 0.014, which is approximately 70\% lower.
\end{itemize}

\begin{table}[!t]
\centering
\caption{Performance results measured in terms of recommendation Top-10 accuracy.}
\label{tbl:results}
\begin{tabular}{llllllllll}
\toprule
\multicolumn{10}{c}{\textbf{Tab 1 - System Role: No (R0), Emphasis: No (E0)}} \\
\toprule
\multicolumn{10}{l}{\textbf{Prompt:} Based on these movies: \{user\_movies\_string\}, recommend 10 movies that the user will likely enjoy.} \\
\multicolumn{10}{l}{\textbf{System Role:} -} \\
\toprule
 &  & \multicolumn{2}{c}{Avg Perf.} & & \multicolumn{2}{c}{Bootstrap Mean} & & \multicolumn{2}{c}{Bootstrap Conf.} \\
\cline{3-4} \cline{6-7} \cline{9-10}Type & Model & NDCG & Recall &  & NDCG & Recall &  & NDCG Conf. & Recall Conf. \\
\toprule
\multirow{6}{*}{\begin{tabular}[c]{@{}l@{}}Generative \\   (ChatGPT)\end{tabular}}   & \textbf{S1.} Simple & \cellcolor{yellow!30}0.008803 & \cellcolor{yellow!30}0.00994 &  & 0.008784 & 0.010011 &  & (0.006323, 0.011484) & (0.00674, 0.013824) \\
& \textbf{S2.} Genre-focused & 0.006964 & 0.007672 &  & 0.006994 & 0.007689 &  & (0.004881, 0.009200) & (0.005079, 0.0106819) \\
& \textbf{S4.} Diversify & 0.001906 & 0.002946 &  & 0.001896 & 0.002946 &  & (0.000822, 0.003185) & (0.001559, 0.004582) \\
& \textbf{S5.} Surprise & 0.007265 & 0.006968 &  & 0.007286 & 0.00696 &  & (0.004671, 0.010279) & (0.004130, 0.010021) \\
& \textbf{S6.} Motivate Reasoning & 0.010152 & 0.010906 &  & 0.010073 & 0.01098 &  & (0.007446, 0.013124) & (0.007511, 0.0148920) \\
& \textbf{S7.} COT & 0.006573 & 0.005848 &  & 0.006554 & 0.005821 &  & (0.004513, 0.008940) & (0.003853, 0.008015) \\
\midrule
\multicolumn{10}{c}{\textbf{Tab 2 - System Role: No (R0), Emphasis: Fair (E1)}} \\
\toprule
\multicolumn{10}{l}{\textbf{Prompt:} Ensure a fair representation of both popular and less-known movies. Based on these movies: \{user\_movies\_string\}, recommend ... .} \\
\multicolumn{10}{l}{\textbf{System Role:} -} \\
\toprule
 &  & \multicolumn{2}{c}{Avg Perf.} & & \multicolumn{2}{c}{Avg. Bootstrap} & \multicolumn{2}{c}{Bootstrap Conf.} \\
\cline{3-4} \cline{6-7} \cline{9-10}Type & Model & NDCG & Recall &  &  NDCG &  Recall &  & NDCG & Recall  \\
\toprule
\multirow{6}{*}{\begin{tabular}[c]{@{}l@{}}Generative \\   (ChatGPT)\end{tabular}}   & \textbf{S1.} Simple & \cellcolor{yellow!30}0.010431 & \cellcolor{yellow!30}0.010557 &  & 0.010404 & 0.010579 &  & (0.007491, 0.013792) & (0.007295, 0.014063) \\
& \textbf{S2.} Genre-focused & 0.005221 & 0.005471 &  & 0.005211 & 0.00555 &  & (0.003413, 0.007300) & (0.003361, 0.008196) \\
& \textbf{S4.} Diversify & 0.001717 & 0.003029 &  & 0.001713 & 0.003024 &  & (0.000607, 0.003261) & (0.001504, 0.005083) \\
& \textbf{S5.} Surprise & 0.004095 & 0.005089 &  & 0.004059 & 0.005112 &  & (0.002399, 0.006132) & (0.002645, 0.008248) \\
& \textbf{S6.} Motivate Reasoning & 0.004315 & 0.005097 &  & 0.004302 & 0.005031 &  & (0.002246, 0.006819) & (0.002668, 0.007909) \\
& \textbf{S7.} COT & 0.007207 & 0.009344 &  & 0.007177 & 0.009373 &  & (0.004493, 0.010162) & (0.005808, 0.013662) \\
\toprule
\multicolumn{10}{c}{\textbf{Tab 3 - System Role: Normal Recommender (R1), Emphasis: Fair (E1)}} \\
\toprule
\multicolumn{10}{l}{\textbf{Prompt:} Based on these movies: \{user\_movies\_string\}, recommend 10 movies that the user will likely enjoy.} \\
\multicolumn{10}{l}{\textbf{System Role:} Act as a fair recommender system balancing between popular and less-known movies to ensure provider fairness.} \\
\toprule
 &  & \multicolumn{2}{c}{Avg Perf.} & & \multicolumn{2}{c}{Avg. Bootstrap} & \multicolumn{2}{c}{Bootstrap Conf.} \\
\cline{3-4} \cline{6-7} \cline{9-10}Type & Model & NDCG & Recall &  & NDCG & Recall &  & NDCG Conf. & Recall Conf. \\
\toprule
\multirow{6}{*}{\begin{tabular}[c]{@{}l@{}}Generative \\   (ChatGPT)\end{tabular}}  & \textbf{S1.} Simple & \cellcolor{green!30}0.013007 & \cellcolor{green!30}0.012606 &  & 0.012984 & 0.012514 &  & (0.009799, 0.016379) & (0.008610, 0.016971) \\
& \textbf{S2.} Genre-focused & 0.00854 & 0.00987 &  & 0.008582 & 0.00983 &  & (0.006277, 0.010984) & (0.006564, 0.013434) \\
& \textbf{S4.} Diversify & 0.00314 & 0.005606 &  & 0.003162 & 0.005523 &  & (0.001477, 0.005548) & (0.003086, 0.008230) \\
& \textbf{S5.} Surprise & 0.006473 & 0.006435 &  & 0.006403 & 0.00642 &  & (0.004065, 0.009319) & (0.003627, 0.009950) \\
& \textbf{S6.} Motivate Reasoning & 0.009379 & 0.00957 &  & 0.009447 & 0.009496 &  & (0.006160, 0.012722) & (0.006056, 0.013061) \\
& \textbf{S7.} COT & 0.013571 & 0.012887 &  & 0.013622 & 0.012941 &  & (0.009836, 0.018061) & (0.009306, 0.017057) \\
\toprule
\multicolumn{10}{c}{\textbf{Tab 4 - System Role: Fair Recommender (R2), Emphasis: No (E0)}} \\
\toprule
\multicolumn{10}{l}{\textbf{Prompt:} \textit{Ensure a fair representation of both popular and less-known movies. Based on these movies: \{user\_movies\_string\}, recommend ... .}} \\
\multicolumn{10}{l}{\textbf{System Role:} \textit{Given a user, act as recommender system.}} \\
\toprule
 &  & \multicolumn{2}{c}{Avg Perf.} & & \multicolumn{2}{c}{Avg. Bootstrap} & \multicolumn{2}{c}{Bootstrap Conf.} \\
\cline{3-4} \cline{6-7} \cline{9-10}Type & Model & NDCG & Recall &  &  NDCG &  Recall &  & NDCG Conf. & Recall Conf. \\
\toprule
\multirow{6}{*}{\begin{tabular}[c]{@{}l@{}}Generative \\   (ChatGPT)\end{tabular}}  & \textbf{S1.} Simple & \cellcolor{green!30}0.014268 & \cellcolor{green!30}0.015796 &  & 0.014257 & 0.015731 &  & (0.011243, 0.017614) & (0.011896, 0.019893) \\
& \textbf{S2.} Genre-focused & 0.008847 & 0.01132 &  & 0.00883 & 0.011382 &  & (0.006700, 0.011233) & (0.007998, 0.015372) \\
& \textbf{S4.} Diversify & \cellcolor{yellow!30}0.002844 & \cellcolor{yellow!30}0.004901 &  & 0.002825 & 0.004872 &  & (0.001303, 0.004922) & (0.003205, 0.006746) \\
& \textbf{S5.} Surprise & 0.007504 & 0.00729 &  & 0.007446 & 0.007304 &  & (0.004587, 0.010622) & (0.004066, 0.011234) \\
& \textbf{S6.} Motivate Reasoning & 0.012354 & 0.011968 &  & 0.012407 & 0.011929 &  & (0.008901, 0.016671) & (0.008133, 0.016052) \\
& \textbf{S7.} COT & 0.013212 & 0.012862 &  & 0.013162 & 0.012908 &  & (0.009972, 0.017185) & (0.009256, 0.016893) \\
\toprule
\multicolumn{10}{c}{\textbf{CF Baselines}} \\
\toprule
\multirow{6}{*}{Discriminative} & BPR-MF & \cellcolor{cyan!30}0.04776 & \cellcolor{cyan!30}0.060699 &  & 0.047742 & 0.060537 &  & (0.040465, 0.054958) & (0.050181, 0.071453) \\
& ItemKNN & \cellcolor{cyan!30}0.04905 & \cellcolor{cyan!30}0.065558 &  & 0.048995 & 0.065367 &  & (0.041122, 0.056398) & (0.054682, 0.076526) \\
& NGCF & 0.04445 & 0.053534 &  & 0.044376 & 0.053448 &  & (0.037306, 0.051784) & (0.044374, 0.063413) \\
& VAE & 0.047158 & 0.056589 &  & 0.047187 & 0.056502 &  & (0.039967, 0.054677) & (0.047093, 0.065803) \\
& LightGCN & 0.048295 & 0.059108 &  & 0.048215 & 0.059019 &  & (0.041621, 0.055491) & (0.049809, 0.068851) \\
& TopPop & 0.030556 & 0.032327 &  & 0.030539 & 0.032127 &  & (0.024253, 0.036734) & (0.025213, 0.039704) \\
\bottomrule
\end{tabular}
\end{table}

\subsubsection*{Answer to Hypotheses} To address the proposed hypotheses, we present the following findings:

\begin{itemize}
    \item \textbf{H1.} \textcolor{cyan}{\textbf{$\checkmark$ Supported}} -- The integration of specific system roles (e.g., \dq{act as a recommender,} or \dq{act as fair recommender}) leads to considerable performance improvements in RecLLMs.
    \item \textbf{H2.} \textcolor{red}{\textbf{$\times$ Rejected}} -- Performance of RecLLMs in zero-shot settings consistently and significantly lags behind CF baselines.
\end{itemize}



\subsubsection{Stability and Confidence of Personalization Metrics}
\label{subsec:stable}
As stated in Section \ref{subsec:rep_exp}, we considered the possibility that external factors such as trending data or algorithmic updates could introduce randomness into responses of GPT-based models. To study the impact of such randomness, each experiment was conducted five times specifically for GPT-based models.\footnote{Note that prior to these experiments, we lacked specific knowledge about the performance of Chat-GPT prompts. Our experiments thus focused on a randomly selected case (\textbf{R0/E0} scenario).} In the analysis of this section, our goal is to observe the variability of GPT-based models across different runs, rather than comparing them to the CF baselines.

\subsubsection*{Hypotheses} To rigorously evaluate the robustness and consistency of GPT-based models under varying conditions, we propose the following hypothesis:

\begin{itemize}
    \item \textbf{H3.} Running multiple iterations of GPT-based recommendation queries for a specific user, results in consistent recommendation quality. 
\end{itemize}

\subsubsection*{Discussion}

\begin{figure}[!h]
    \centering
    \begin{subfigure}[b]{0.95\textwidth}
        \includegraphics[width=\textwidth]{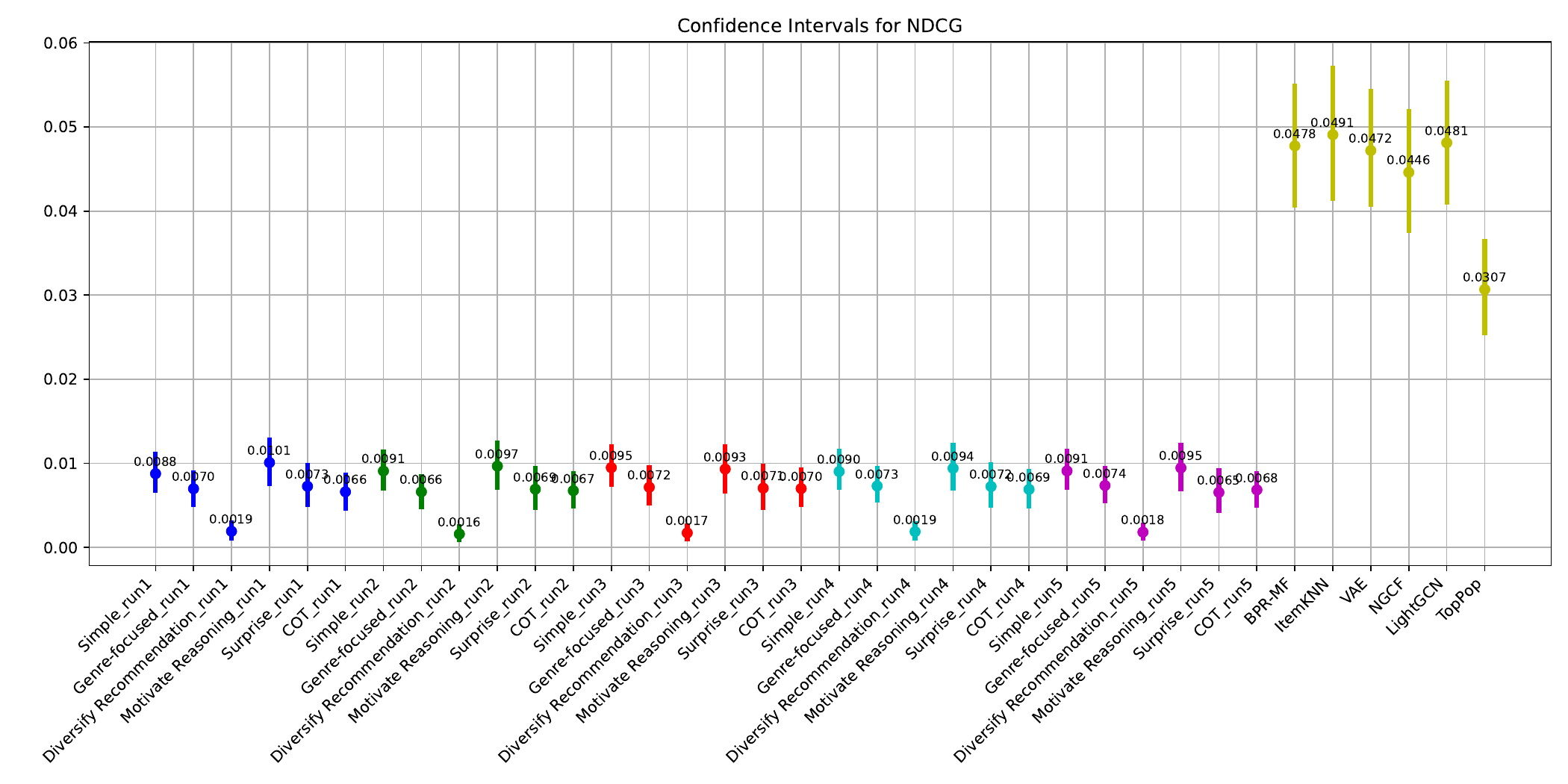}
        \caption{NDCG@10.}
    \end{subfigure}
    \\
    \begin{subfigure}[b]{0.95\textwidth}
        \includegraphics[width=\textwidth]{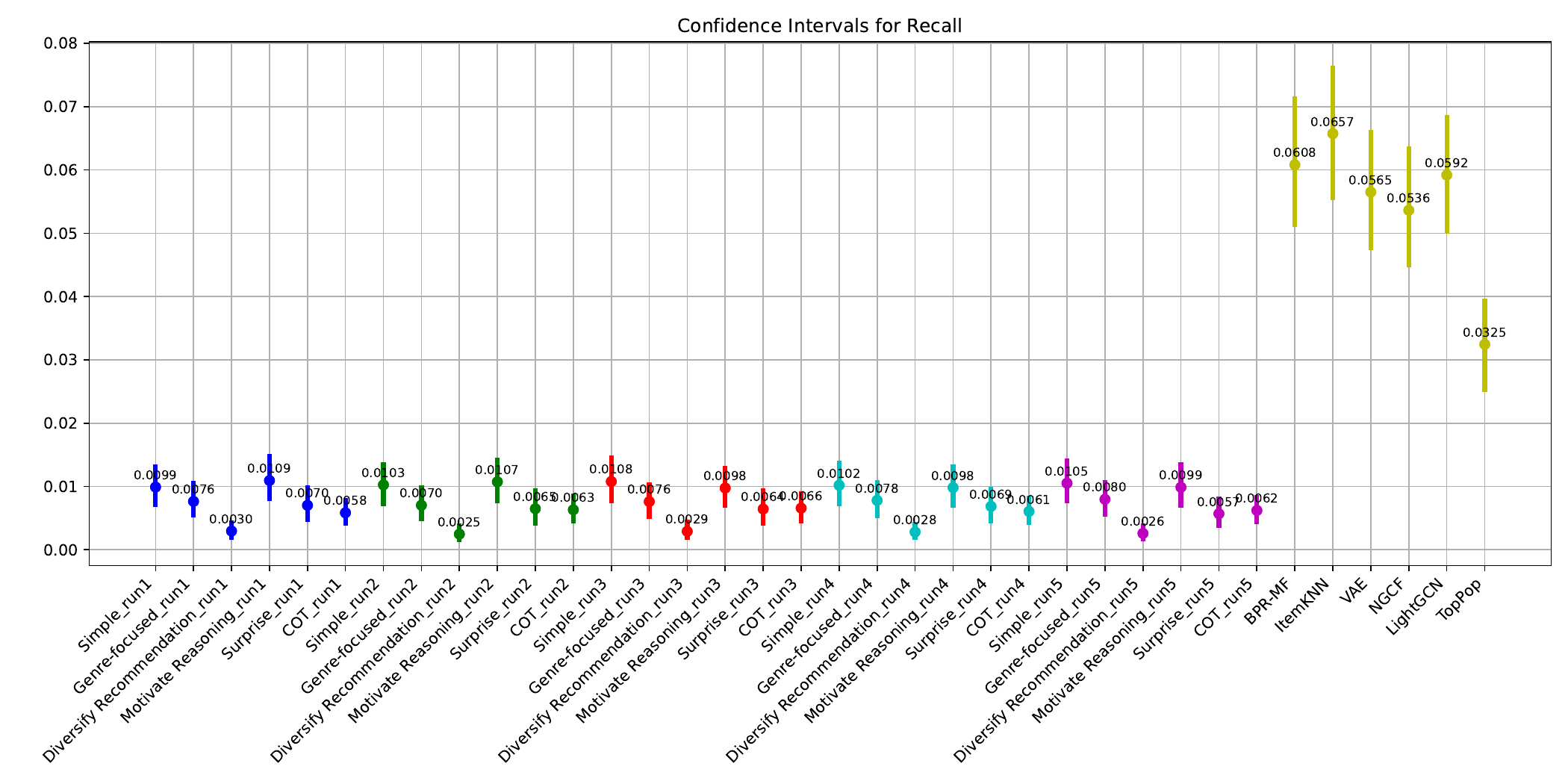}
        \caption{Recall@10}
    \end{subfigure}
    \caption{Studying the stability of GPT-based performance metric across different runs.\label{fig:runs}
}
\end{figure}

The graph in Figures~\ref{fig:runs} illustrates the NDCG and Recall scores for GPT-based recommendation models across five different runs. These runs overall have been conducted at different points in time that could span from \textit{hours to a few days}. Notably, there is observable variability in the performance metrics from run to run. For instance, 
\begin{itemize}
 \item \textbf{Simple} (\textbf{S1}): The NDCG scores are [0.0088, 0.0091, 0.0095, 0.0094, 0.0095], with a std\footnote{standard deviation} of approximately 0.000305.
\item \textbf{Motivate reasoning} (\textbf{S6}): The NDCG scores are [0.0101, 0.0097, 0.0093, 0.0094, 0.0095], with a std of approximately 0.000316.
\item \textbf{COT} (\textbf{S7}): The NDCG scores are [0.0066, 0.0067, 0.0070, 0.0069, 0.0068], with a std of approximately 0.000158.
\end{itemize}

Despite the inherent randomness in individual iterations, the general tendency of the results remains consistent, suggesting that while individual runs may produce fluctuating scores, the overall behavior of the models does not significantly deviate. This is evidenced by the low standard deviations in the NDCG scores (0.000305 for \textbf{S1}, 0.000316 for \textbf{S6}, and 0.000158 for \textbf{S7}), indicating a high level of result stability over different runs. Such consistency aligns with the expected behavior of generative models, which can exhibit some degree of randomness in their output due to the stochastic nature of the underlying data/algorithms. However, the close clustering of these scores, as reflected in the narrow stds, indicates a stable pattern of performance.

While the current study demonstrates a high degree of stability in model results over multiple runs, it is crucial to note that these findings do not necessarily guarantee stability over longer time frames, such as weeks or months. The dynamic nature of data and algorithms in generative models means that updates and changes over time could lead to different behaviors and outputs. Nonetheless, \textit{caution is required} when generalizing these results, and continuous monitoring and periodic re-evaluation of the performance over extended periods are recommended to ensure that the insights remain relevant and accurate. 

\subsubsection*{Answer to hypotheses} To evaluate the validity of our proposed hypotheses, we present detailed findings below, supported by statistical data and analyses:

\begin{itemize}
    \item \textbf{H3.} \textcolor{cyan}{\textbf{$\checkmark$ Supported}} -- This hypothesis, which anticipated consistent performance across different GPT-based model runs, is supported. The data presented in the NDCG graph aligns with this expectation, showing only minor variability across multiple runs. Confidence intervals overlap significantly for different runs of each GPT-based model, which confirms the reliability of the system in providing personalized recommendations. However, it is important to consider these results with caution for long-term deployment as they do not necessarily guarantee stability over extended periods such as weeks or months.
\end{itemize}

\subsubsection{Item Fairness}
\label{subsec:prov_fairness}

\bluegreen{This section explores the item fairness metrics of different recommender models, crucial for evaluating their effectiveness in providing balanced and varied recommendations.

\subsubsection*{Hypotheses} To ascertain the impact of different model strategies on recommendation fairness, we carefully analyzed the outcomes in the context of our formulated hypotheses:
\begin{itemize}
  \item \textbf{H4:} Due to their training on large-scale internet data, GPT-based models are hypothesized to exhibit lower item fairness than classical CF models. However, asking the system to produce more diverse or additional recommendations (scenarios \textbf{S3} and \textbf{S4}) is expected to increase/improve item fairness.
  \item \textbf{H5:} Integrating a \dq{Fair Recommender} as a system role in GPT-based systems is expected to enhance item fairness.
\end{itemize}

\begin{table}[!t]
\caption{Provider fairness of the tested models in Experiment 1}
\label{subsec:div}
\begin{tabularx}{\textwidth}{lcccccc}
\toprule
\textbf{Model} & \multicolumn{3}{c}{\textbf{Normal Recommender}} & \multicolumn{3}{c}{\textbf{Fair Recommender}} \\
\cmidrule(r){2-4} \cmidrule(l){5-7}
& \textbf{Gini Coefficient $\downarrow$} & \textbf{HHI $\downarrow$} & \textbf{Entropy$\uparrow$} & \textbf{Gini Coefficient $\downarrow$} & \textbf{HHI $\downarrow$} & \textbf{Entropy$\uparrow$}\\
\midrule
Simple & \cellcolor{yellow!30}0.982463 & 0.017204 & \cellcolor{yellow!30}5.042821 & \cellcolor{green!30}0.978925 & 0.010899 & \cellcolor{green!30}5.387465 \\
Genre-focused & \cellcolor{green!30}0.964743 & 0.006455 & \cellcolor{green!30}5.919697 & \cellcolor{green!30}0.959879 & 0.004771 & \cellcolor{green!30}6.110040 \\
Diversify Recommendation & \cellcolor{yellow!30}0.992349 & 0.034724 & \cellcolor{yellow!30}4.232139 & \cellcolor{yellow!30}0.992603 & 0.030010 & \cellcolor{yellow!30}4.321307 \\
Surprise & \cellcolor{yellow!30}0.997906 & 0.059857 & \cellcolor{yellow!30}3.227737 & \cellcolor{yellow!30}0.998365 & 0.067952 & \cellcolor{yellow!30}3.023948 \\
Motivate Reasoning & \cellcolor{green!30}0.981745 & 0.019189 & \cellcolor{green!30}5.026322 & \cellcolor{green!30}0.979133 & 0.011218 & \cellcolor{green!30}5.366627 \\
Chain-of-thought (COT) & \cellcolor{green!30}0.986889 & 0.027030 & \cellcolor{green!30}4.619500 & \cellcolor{green!30}0.979313 & 0.011167 & \cellcolor{green!30}5.365294 \\ \midrule
BPR-MF & \cellcolor{yellow!30}0.991758 & 0.012550 & \cellcolor{yellow!30}4.658056 & \cellcolor{yellow!30}0.991758 & 0.012550 & \cellcolor{yellow!30}4.658056 \\
Item-KNN & \cellcolor{cyan!30}0.914271 & 0.002877 & \cellcolor{cyan!30}6.671847 & \cellcolor{cyan!30}0.914271 & 0.002877 & \cellcolor{cyan!30}6.671847 \\
NGCF & \cellcolor{cyan!30}0.950845 & 0.002762 & \cellcolor{cyan!30}6.420996 & \cellcolor{cyan!30}0.950845 & 0.002762 & \cellcolor{cyan!30}6.420996 \\
VAE & \cellcolor{yellow!30}0.989722 & 0.009554 & \cellcolor{yellow!30}4.903511 & \cellcolor{yellow!30}0.989722 & 0.009554 & \cellcolor{yellow!30}4.903511 \\
LightGCN & \cellcolor{yellow!30}0.989610 & 0.010546 & \cellcolor{yellow!30}4.861879 & \cellcolor{yellow!30}0.989610 & 0.010546 & \cellcolor{yellow!30}4.861879 \\
TopPop & \cellcolor{yellow!30}0.994859 & 0.020000 & \cellcolor{yellow!30}3.912023 & \cellcolor{yellow!30}0.994859 & 0.020000 & \cellcolor{yellow!30}3.912023 \\
\bottomrule
\end{tabularx}
\begin{tabular}{l l l}
  \cellcolor{cyan!30} & Cyan shows the best performing methods. &\\
  \cellcolor{green!30} & Green shows good performing methods (relative to others). &\\
  \cellcolor{yellow!30} & Yellow shows lower performance models.&
\end{tabular}
\end{table}

\subsubsection*{Results and Discussion} 

We now address the above research hypotheses, using Table~\ref{subsec:div} to inform our discussion. We primarily use the Gini coefficient and entropy for this analysis, noting that HHI shows somewhat similar performance to the Gini coefficient. Overall, by examining the Gini indices and entropy values, we observe that two CF models, Item-KNN and NGCF, provide the best performances in terms of item fairness. Among the RecLLM models, the \textit{Genre-focused,} \textit{Motivated Reasoning,} and \textit{Chain-of-thought} scenarios exhibit notable item fairness after the CF baselines.

Summarizing the best-performing models, we have: $ItemKNN/NGCF > Genre/Motivate/COT > rest$. One notable and interesting result from the table is that \textit{asking GPT to produce novel or diverse recommendations has little or no significant impact on improving its recommendability}. The entropy generally increases for RecLLM models when switching to Fair recommenders, indicating higher novelty in recommendations. This pattern is not observed in the CF models, which show no change in their metrics. To address~\textbf{H4}, we find that GPT-based models generally exhibit lower item fairness compared to classical CF models, as hypothesized. However, contrary to the expectation, asking the system to produce more diverse or additional recommendations did not significantly increase item fairness. For~\textbf{H5}, integrating a \dq{Fair Recommender} role in GPT-based systems did enhance recommendation fairness, as indicated by the increased entropy in the RecLLM models. This suggests a mitigation of item unfairness and an enhancement of recommendation diversity.

\subsubsection*{Answer to Hypotheses}
We can answer to the above hypotheses as following:

\begin{itemize}
  \item \textbf{H4:} \textcolor{orange}{\textbf{$\circ$ Partially Supported}} — While classical CF models such as Item-KNN and NGCF exhibit higher fairness compared to RecLLM models, the GPT-based models named \textit{"Genre-focused"} and \textit{"Motivated Reasoning"} showed improvements in fairness. This hypothesis is partially supported as diversification strategies and requests for more novel recommendations did not increase item fairness in RecLLM models.
  
  \item \textbf{H5:} \textcolor{cyan}{\textbf{$\checkmark$ Supported}} — Implementing a \textit{"Fair Recommender"} role within GPT-based systems leads to noticeable improvements in fairness. This supports the hypothesis that transitioning from normal to fair recommender systems enhances fairness, demonstrating the controllability in RecLLM models, particularly in terms of diversity and novelty.
\end{itemize}}



\subsubsection{Analysis of Temporal Freshness and Genre Bias in Recommender Systems}
\label{subsec_temporalFesh}

In this unified analysis, we explore the tendencies of various GPT-based and CF models in terms of recency and genre preferences in movie recommendations. Our study aims to discern whether these models demonstrate a bias towards recommending newer or older films and if they exhibit specific genre preferences that could illuminate performance differences between model types.

\subsubsection*{Research Hypotheses}
\bluegreen{We provide the following statement of research hypotheses:
\begin{itemize}
\item \textbf{H6.} We hypothesize that GPT-based models, given their extensive training on diverse and recent datasets, are more likely to recommend \textbf{newer films} compared to classical CF models, which may favor older, well-established films.
\item \textbf{H7.} We expect that GPT-based models will show a broader genre distribution in their recommendations due to their training on diverse data sources, compared to CF models, which may adhere more strictly to users' historical genre preferences.
\end{itemize}}

\begin{figure}[!h]
    \centering
    \begin{minipage}{0.65\textwidth}
        \centering
        \includegraphics[width=\textwidth]{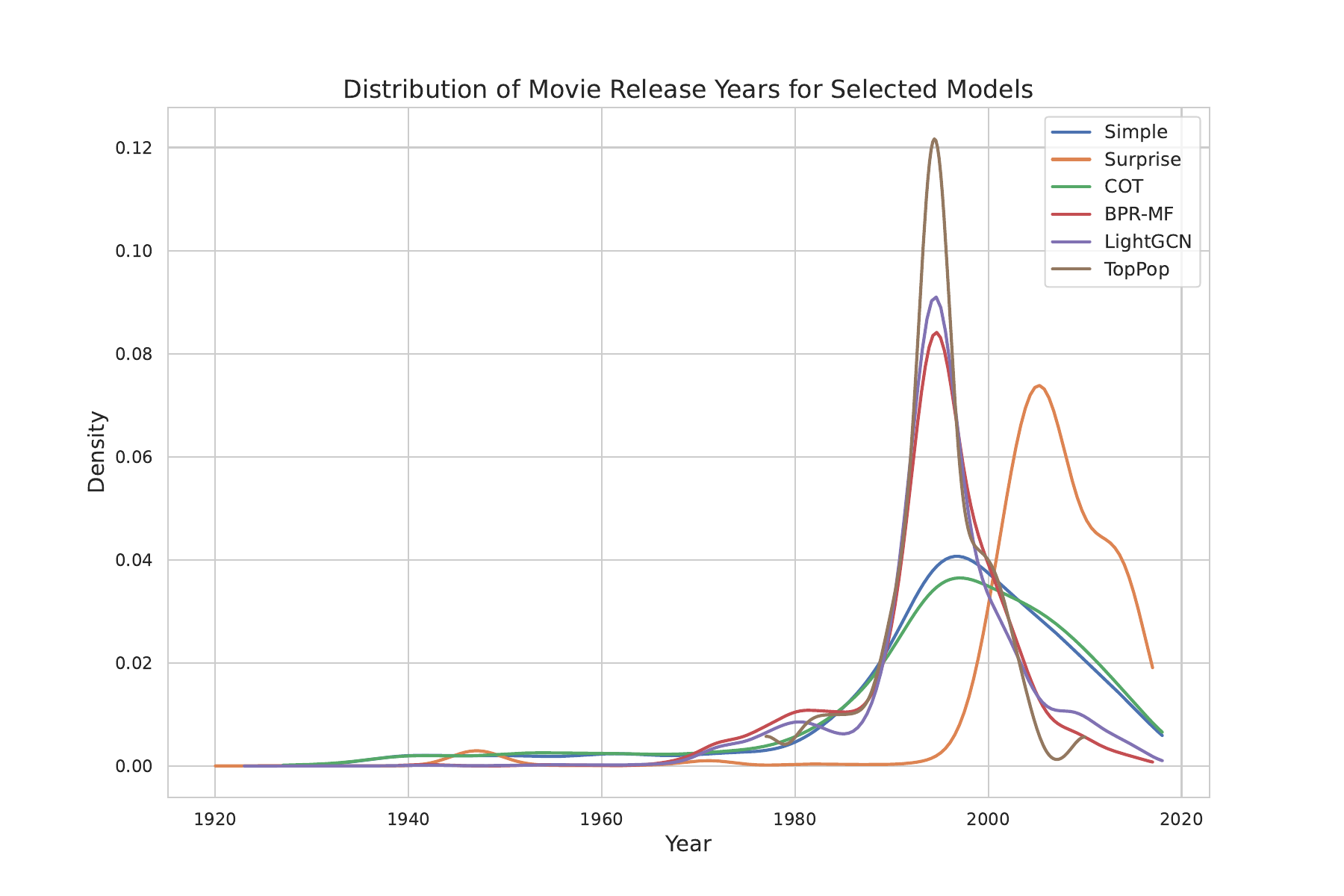} 
        \caption{Distribution of Movie Release Years as recommended by different models}
        \label{fig:movie_years}
    \end{minipage}\hfill
    \begin{minipage}{0.34\textwidth}
        \centering
        \captionof{table}{Model Statistics Reflecting Recency Bias}
        \label{tab:model_stats}
        \begin{tabular}{lrr}
            \toprule
            \textbf{Model}  & \textbf{Median Year} & \textbf{Std Year} \\
            \toprule
            \textbf{Simple}  & 1999 & 15.09 \\
            \textbf{Genre-focused } & 1997 & 14.99 \\
            \textbf{Diversify}  & 2007 & 11.76 \\
            \textbf{Surprise}  & 2006 & 10.62 \\
            \textbf{Motivate reasoning}  & 2002 & 15.71 \\
            \textbf{COT}  & 1999 & 15.82 \\
            \midrule
            \textbf{BPR-MF}  & 1995 & 8.27 \\
            \textbf{ItemKNN}  & 1995 & 12.14 \\
            \textbf{NGCF}  & 1995 & 12.25 \\
            \textbf{VAE}  & 1995 & 8.50 \\
            \textbf{LightGCN}  & 1995 & 8.34 \\
            \textbf{Pop} & 1995 & 5.65 \\
            \bottomrule
        \end{tabular}
    \end{minipage}
\end{figure}



\bluegreen{\subsubsection*{Results and Discussions} Table \ref{tab:model_stats} presents the statistics that inform our understanding of each model tendency towards recommending newer or older films. Notably, the GPT-based models \sq{Surprise} and \sq{Diversify} push the boundaries towards recommending more recent items, with median release years of 2006 and 2007. This indicates a clear orientation towards newer movies, possibly interpreting \sq{surprise} and \sq{diversification} in their temporal sense. Since GPT models do not possess inherent information about the novelty of content in terms of user interactions, they likely interpret these directives based on the freshness of the content itself. In contrast, the CF models tend to recommend older films, median release years consistently at 1995, reflecting a stronger inclination towards well-established, historically popular films. This distinction suggests that GPT-based models might be leveraging their extensive training datasets to prioritize more recent films, which can be beneficial in scenarios where up-to-date recommendations are desired. The broader temporal range of recommendations from GPT-based models, as evidenced by the standard deviation in the year of movies recommended (Surprise: Std Year 10.62, Diversify: Std Year 11.76), further underscores their flexibility in catering to diverse temporal tastes.

Upon analyzing the word clouds, it is evident that CF models such as \textbf{BPR-MF}, \textbf{LightGCN}, and \textbf{VAE} predominantly recommend genres such as \dq{Action,} \dq{Adventure,} \dq{Thriller,} and \dq{Sci-Fi.} These genres appear frequently and in larger font sizes, indicating their dominance in the recommendations made by these models. In contrast, GPT-based models, represented in the lower word clouds, show a more varied genre distribution with \dq{Comedy,} \dq{Drama,} and \dq{Romance} and even \dq{Thriller} and \dq{Adventure} being more prominent. These results are interesting and suggest that GPT-based RecLLMs may be less constrained by the users' historical genre affinities and more exploratory in their recommendations, possibly leading to a broader but potentially less precise genre match to individual user profiles.

\begin{figure}[!b]
    \centering
    \includegraphics[width=0.850\textwidth]{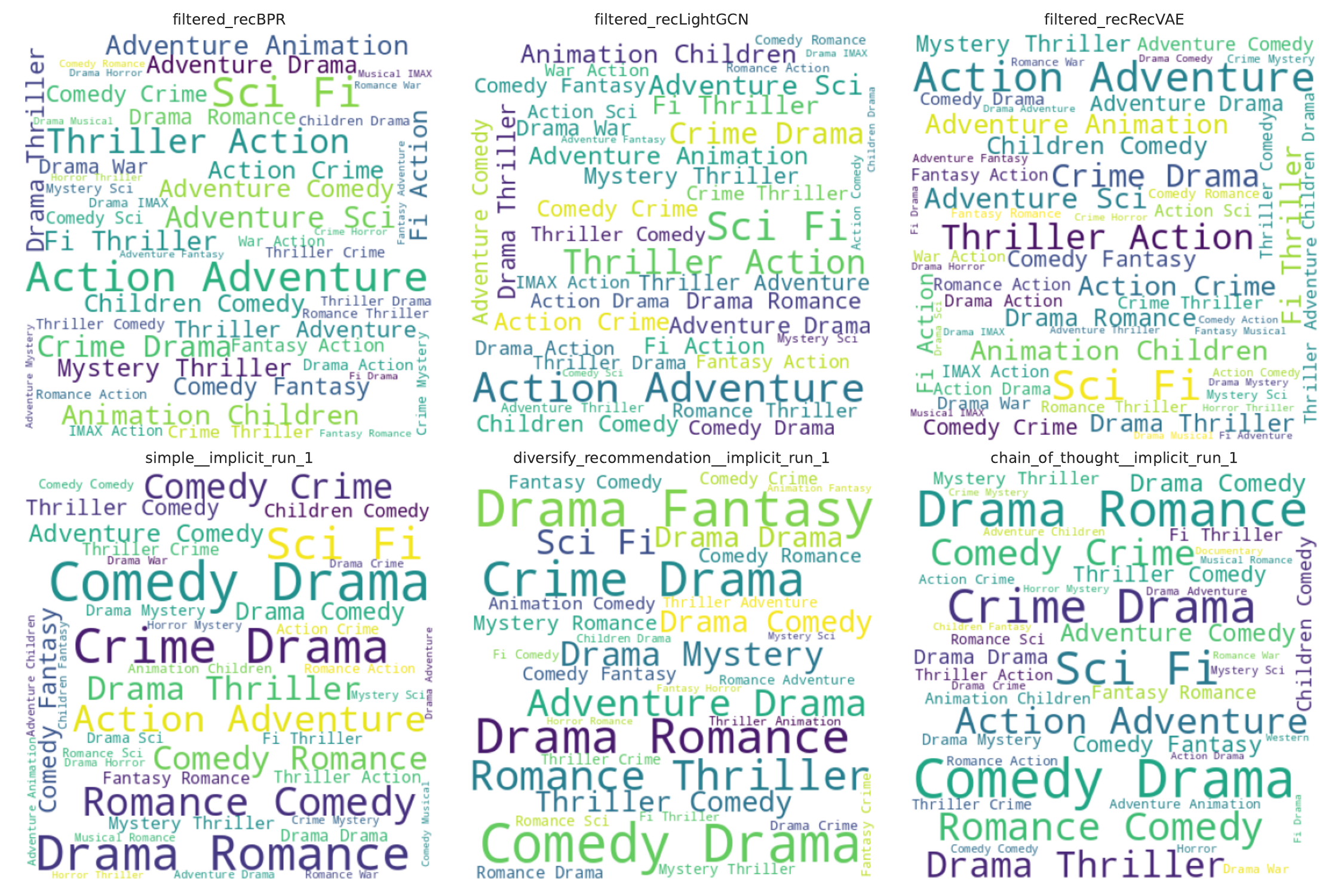} 
    \caption{WordCloud of Movie Genres as recommended by different models. Top model correspond to CF models (\textbf{BPR-MF}, \textbf{LightGCN}, \textbf{RecVAE}), while  lower models include GPT-based recommenders (\textbf{Simple}, \textbf{Diversity}, \textbf{COT}).)}
    \label{fig:movie_years}
\end{figure}

\begin{itemize}
    \item \textbf{H6.} \textcolor{cyan}{\textbf{$\checkmark$ Supported}} -- GPT-based models, particularly \textit{Diversify} and \textit{Surprise}, tend to recommend significantly newer movies than CF models. This finding supports the hypothesis, confirming that GPT-based models are effective in capturing recent trends and offering more temporally relevant recommendations.
    \item \textbf{H7.} \textcolor{orange}{\textbf{$\circ$ Partially Supported}} -- While CF models are found to predominantly recommend genres such as \textit{Action} and \textit{Sci-Fi}, GPT-based models demonstrate a broader genre distribution, including \textit{Comedy}, \textit{Drama}, and \textit{Romance}. This diversity in GPT-based recommendations supports the hypothesis to a certain extent, indicating that these models are utilizing their extensive training on diverse datasets to introduce a wider variety of content.
\end{itemize}}

\noindent \textbf{Implicit vs. explicit scenario.} Analyzing the impact of including ratings in the prompt shows noteworthy results. In our study, we initially avoided using ratings for the GPT models, aligning with the implicit setting of the Collaborative Filtering (CF) baseline. However, for completeness, we explored the effect of revealing user ratings for selected items. 

The \dq{Include-rating} model demonstrates a significant improvement compared to other GPT-based scenarios. With an NDCG of 0.013230 and a Recall of 0.014721, it surpasses genre-focused, Diversify Recommendation, Surprise, Motivate Reasoning, and chain-of-thought (COT) models. The Bootstrap NDCG and Recall for the Include-rating model are 0.013280 and 0.014673, respectively, indicating consistency in performance. The confidence intervals for NDCG (0.009737 to 0.017012) and Recall (0.010261 to 0.019795) further emphasize its reliability.

These results suggest that including ratings significantly enhances the recommendation performance of the system. It provides a more personalized and accurate reflection of user preferences, which is a key aspect in recommendation systems. While other GPT-based models showed varied performance, the inclusion of ratings seems to offer a more consistent and improved outcome. This insight could be valuable in refining other scenarios and models, potentially leading to better overall system performance.

\subsection{Sequential Recommendations}
\label{subsec:seq_rec}

\bluegreen{This section focuses on the performance of RecLLMs in sequential recommendation task, measured by recommendation top-$K$ accuracy, item fairness, and coverage. Results are presented in Table~\ref{tab:sim_align}, with each generative scenario (using ChatGPT) in the upper sections and CF baselines in the bottom section.

\subsubsection*{Hypotheses} To investigate the effectiveness of different recommendation strategies, we propose the following hypotheses:

\begin{itemize}
    \item \textbf{H1.} Incorporating sequential in-context learning (ICL) prompts, including \textit{ICL-1} and \textit{ICL-2}, results in improved recommendation accuracy compared to zero-shot learning.
    \item \textbf{H2.} Sequential in-context learning (ICL) prompts will result in better item fairness and coverage compared to zero-shot learning.
    \item \textbf{H3.} A combination of improved accuracy, item fairness, and coverage will be observed in certain settings, highlighting the optimal recommendation strategy.
\end{itemize}

\subsubsection*{Discussion.}

Our evaluation highlights several findings regarding the personalization of recommendations using RecLLM.

\begin{itemize}
    \item Regarding \textbf{H1}, the results indicate that recommendation \textbf{accuracy,} measured by Hit Rate (HR) and NDCG, varies with different learning strategies. For the \texttt{MovieLens} dataset, \textit{without demographic information,} the HR for zero-shot learning is 0.179, which decreases slightly for ICL-1 (0.154) and ICL-2 (0.150). In the \texttt{LastFM} dataset, zero-shot learning HR is 0.204, which decreases to 0.154 for both ICL-1 and ICL-2. 

    However, \textit{including demographic information,} especially in the \texttt{Age-Group} category, improves accuracy in the movie domain. For instance, in the \texttt{MovieLens} dataset with age-group context, the HR for ICL-2 improves to 0.204, which is higher than zero-shot learning with such information (0.171) or zero-shot learning with no demographic information (0.179). Another clear trend is the impact of the Recency-Focused (Rec.-Freq) sampling in the context part on recommendation accuracy. When sensitive attributes are included, Rec-Freq and ICL-2 generally show better performance compared to the same age-category included in zero-shot learning. For example:
   \vspace{1mm}
    \begin{itemize}
        \item In the \texttt{MovieLens} dataset with age-group context, the HR for Rec-Freq ICL-2 is 0.204, compared to 0.171 for zero-shot learning.
        \item In the \texttt{MovieLens} dataset with gender context, the HR for Rec-Freq ICL-2 is 0.175, compared to 0.158 for zero-shot learning.
        \item In the \texttt{LastFM} dataset with age-group context, the HR for Rec-Freq ICL-2 is 0.204, compared to 0.263 for zero-shot learning.
        \item In the \texttt{LastFM} dataset with gender context, the HR for Rec-Freq ICL-2 is 0.204, compared to 0.221 for zero-shot learning.
    \end{itemize}
    \vspace{1.5mm}
    In terms of NDCG, however, results generally favor zero-shot learning scenarios across different combinations of demographic information and sampling strategies in both datasets. While there is a slight improvement in NDCG for the Rec-Freq ICL-2 scenario in the \texttt{LastFM} dataset compared to other scenarios (zero-shot or ICL-1), the differences are marginal. For example:

    \begin{itemize}
        \item In the \texttt{LastFM} dataset with age-group context, the NDCG for Rec-Freq ICL-2 is 0.60000, compared to 0.56250 for ICL-1, and 0.82500 for zero-shot learning.
        \item In the \texttt{LastFM} dataset with gender context, the NDCG for Rec-Freq ICL-2 is 0.65625, compared to 0.59375 for ICL-1, and 0.64375 for zero-shot learning.
    \end{itemize}

    These examples illustrate that incorporating sensitive attributes and employing the Rec-Freq sampling method in ICL-2 can enhance recommendation accuracy, although the improvements in NDCG are not as pronounced.
    \vspace{2mm}
     
    \item Regarding \textbf{H2}, \textbf{item fairness and coverage} show mixed results across different settings. For item fairness, as measured by Gini and Entropy, zero-shot learning and ICL strategies show slight differences. For instance, in the \texttt{MovieLens} dataset with no demographic info, Gini for zero-shot learning is 0.57739 (lower is better), which increases to 0.66303 for ICL-1 and 0.65835 for ICL-2, indicating worse fairness. Entropy, where higher values are better, decreases from 3.47952 (zero-shot) to 2.87493 (ICL-1) and 2.88530 (ICL-2), suggesting that ICL prompts might lead to slightly worse item fairness. \textbf{Coverage} remains consistently low across all scenarios in the movie domain, and zero-shot learning provides considerably better coverage performance. For example:

    \begin{itemize}
        \item In the \texttt{MovieLens} dataset without demographic information, the coverage for zero-shot learning is 0.02621, compared to 0.01685 for ICL-1 and ICL-2.
        \item In the \texttt{MovieLens} dataset with gender context, the coverage for zero-shot learning is 0.02845, compared to 0.01498 for ICL-2.
    \end{itemize}

    In the music domain, results generally favor ICL-2. For example:

    \begin{itemize}
        \item In the \texttt{LastFM} dataset with age-group context, the coverage for Rec-Freq ICL-2 is 0.00210, compared to 0.00198 for zero-shot learning.
        \item In the \texttt{LastFM} dataset with gender context, the coverage for Rec-Freq ICL-2 is 0.00218, compared to 0.00211 for zero-shot learning.
    \end{itemize}

    Notably, including demographic information generally enhances fairness metrics. For example, with \texttt{Gender} context in \texttt{MovieLens}, the Gini index improves, and Entropy increases, indicating fairer item distribution.

    \item Regarding \textbf{H3}, a combination of improved accuracy, item fairness, and coverage is generally observed with the \texttt{Recency-Focused} approach in both zero-shot and ICL-2 scenarios. By visually examining the placements of highlighted values, we note that \texttt{Recency-Focused} zero-shot and ICL-2 tend to provide the best combined performance of accuracy, item fairness, and coverage. Additionally, revealing demographic information, particularly age-group (young/old) and gender, has a noticeable positive impact on the fairness and coverage of the methods, thereby enhancing the utility of the system from both consumer and producer perspective.

    \end{itemize}

\begin{table}[!t]
\caption{Recommendation alignment between  ($\mathcal{R}_{m}$, ~$\mathcal{R}_{m}^a$) based on Item Similarity \(\beta_{item}\)}
\label{tab:sim_align}
\centering
\renewcommand{\arraystretch}{1.1} 
\resizebox{\textwidth}{!}{%
\begin{tabular}{lccccccccccccccccc}
\toprule
\small{\textbf{Attribute}} &
\textbf{Type} &
\textbf{ICL} &
\multicolumn{7}{c}{\texttt{MovieLens}} & & 
\multicolumn{7}{c}{\texttt{LastFM}} \\
\cline{4-10} \cline{12-18}
& & & \multicolumn{2}{c}{\textbf{Acc.}}  & & \multicolumn{2}{c}{\textbf{Item Fairness}} & & \textbf{Cov.} & & 
\multicolumn{2}{c}{\textbf{Acc.}} & & \multicolumn{2}{c}{\textbf{Item Fairness}} & & \textbf{Cov.} \\
\cline{4-5} \cline{7-8} \cline{10-10} \cline{12-13} \cline{15-16} \cline{18-18}

& & & \small{HR} & \small{NDCG} & & \small{Gini} & \small{Entropy} & & - & & \small{HR}  & \small{NDCG}  &  &  \small{Gini} & \small{Entropy} & & - \\
\midrule
\multirow{6}{*}{\textbf{\color{blue}{No Info}}} & \multirow{3}{*}{\textbf{Freq.}} & 0-shot & \cellcolor{green!30}0.179 & 0.53750 & & \cellcolor{yellow!30}0.57739 & \cellcolor{yellow!30}3.47952 & &  \cellcolor{green!30}0.02621 & & \cellcolor{green!30}0.204 & 0.66875 & & 0.24882 & 4.89089 & & 0.00195 \\
 &  & ICL-1 &  \cellcolor{yellow!30}0.154 & \cellcolor{yellow!30}0.55000 & & \cellcolor{yellow!30}0.66303 & \cellcolor{yellow!30}2.87493 & & \cellcolor{yellow!30}0.01685 & & \cellcolor{yellow!30}0.154 & 0.47500 & & 0.24464 & 4.91536 & & 0.00201 \\
 &  & ICL-2 &  \cellcolor{yellow!30}0.150 & 0.40000 & & \cellcolor{yellow!30}0.65835 & \cellcolor{yellow!30}2.88530 & & \cellcolor{yellow!30}0.01685 & & \cellcolor{yellow!30}0.154 & 0.44375 & & 0.25137 & 4.92419 & & 0.00201 \\
\cline{2-18}
 & \multirow{3}{*}{\textbf{Rec.-Freq.}} & 0-shot & 0.208 & 0.65000 & & 0.55075 & 3.66855 & & 0.03033 & & 0.225 & 0.65000 & & 0.18229 & 5.07136 & & 0.00220 \\
 &  & ICL-1 & 0.162 & 0.54375 & & 0.60812 & 3.08940 & & 0.01760 & & 0.183 & 0.53750 & & 0.20217 & 4.99446 & & 0.00206 \\
 &  & ICL-2 & 0.175 & 0.56875 & & 0.64473 & 2.92146 & & 0.01610 & & 0.208 & 0.58125 & & 0.20439 & 5.00125 & & 0.00205 \\
\cline{1-18}
\multirow{6}{*}{\textbf{\color{red}{Gender}}} & \multirow{3}{*}{\textbf{Freq.}} & 0-shot & 0.146 & \cellcolor{green!30}0.58750 & & 0.56064 & 3.60813 & &  \cellcolor{green!30}0.02845 & & 0.196 & 0.68750 & & 0.21299 & 5.00305 & & 0.00211 \\
 &  & ICL-1 &  0.167 & 0.50625 & & 0.67541 & 2.71790 & & \cellcolor{yellow!30}0.01460 & & 0.158 & 0.51875 & & 0.26066 & 4.82855 & & 0.00188 \\
 &  & ICL-2 &  0.171 & 0.43125 & & 0.66145 & 2.76958 & & \cellcolor{yellow!30}0.01498 & & 0.154 & 0.42500 & & 0.22654 & 4.99862 & & 0.00215 \\
\cline{2-18}
 & \multirow{3}{*}{\textbf{Rec.-Freq.}} & 0-shot & \cellcolor{yellow!30}0.158 & 0.55000 & & 0.54779 & 3.68560 & & 0.02958 & & \cellcolor{green!30}0.221 & \cellcolor{yellow!30}0.64375 & & 0.18254 & 5.08343 & & \cellcolor{cyan!30}0.00223 \\
 &  & ICL-1 & 0.154 & 0.46250 & & 0.64715 & 2.99814 & & 0.01909 & & 0.208 & \cellcolor{yellow!30}0.59375 & & 0.19480 & 5.01322 & & 0.00209 \\
 &  & ICL-2 & \cellcolor{green!30}0.175 & 0.50625 & & 0.64839 & 2.73183 & & 0.01310 & & \cellcolor{yellow!30}0.204 & \cellcolor{green!30}0.65625 & & 0.18622 & 5.06197 & & \cellcolor{green!30}0.00218 \\
\cline{1-18}
\multirow{6}{*}{\textbf{\color{green}{Age-Group}}} & \multirow{3}{*}{\textbf{Freq.}} & 0-shot & 0.154 & 0.70000 & & 0.54193 & 3.72305 & &  \cellcolor{cyan!30}0.03145 & & 0.183 & 0.56875 & & 0.24498 & 4.91475 & & 0.00198 \\
 &  & ICL-1 &  0.175 & 0.57500 & & 0.67186 & 2.73694 & & 0.01498 & & 0.146 & 0.48750 & & 0.24892 & 4.90102 & & 0.00199 \\
 &  & ICL-2 &  0.150 & 0.45625 & & 0.64637 & 2.87851 & & 0.01572 & & 0.146 & 0.41250 & & 0.23557 & 4.97022 & & 0.00211 \\
\cline{2-18}
 & \multirow{3}{*}{\textbf{Rec.-Freq.}} & 0-shot & \cellcolor{yellow!30}0.171 & 0.66875 & & 0.56433 & 3.63473 & & 0.02995 & & \cellcolor{cyan!30}0.263 & \cellcolor{cyan!30}0.82500 & & 0.18816 & 5.07481 & & 0.00223 \\
 &  & ICL-1 & 0.158 & 0.46875 & & 0.63580 & 2.92408 & & 0.01610 & & 0.188 & \cellcolor{yellow!30}0.56250 & & 0.20575 & 4.96456 & & 0.00200 \\
 &  & ICL-2 &  \cellcolor{cyan!30}0.204 & 0.60000 & & 0.63038 & 2.93595 & & 0.01498 & & \cellcolor{yellow!30}0.204 & \cellcolor{green!30}0.60000 & & 0.19985 & 5.01733 & & 0.00210 \\
\bottomrule
\end{tabular}
}
\vspace{0.3cm}
\begin{tabular}{l l l}
  \cellcolor{cyan!30} & Cyan shows the best performing methods. &\\
  \cellcolor{green!30} & Green shows good performing methods (relative to others). &\\
  \cellcolor{yellow!30} & Yellow shows lower performance models.&
\end{tabular}
\end{table}

\subsubsection*{Answer to Hypotheses} To address the proposed hypotheses, we present the following findings:

\begin{itemize}
    \item \textbf{H1.} \textcolor{orange}{\textbf{$\circ$ Partially Supported}} -- The incorporation of sequential in-context learning prompts (ICL-1 and ICL-2) improves recommendation accuracy in certain contexts, particularly with demographic information (e.g., \texttt{Age-Group} in \texttt{MovieLens}), though not consistently across all settings.
    \item \textbf{H2.} \textcolor{red}{\textbf{$\times$ Rejected}} -- Sequential in-context learning prompts do not consistently lead to better item fairness and coverage compared to zero-shot learning. In some cases, they worsen item fairness, as indicated by higher Gini indices and lower Entropy values.
    \item \textbf{H3.} \textcolor{cyan}{\textbf{$\checkmark$ Supported}} -- Certain settings, particularly \texttt{Recency-Focused ICL-2} with demographic information, show a balance between accuracy and item fairness, indicating potential optimal recommendation strategies.
\end{itemize}}

\bluegreen{\subsection{Economic and Practical Implications}\label{subsec:scaliabity}

Integrating large language models (LLMs) like GPT-3.5-turbo into recommender systems involves several economic and practical considerations. Below, we discuss the inference costs, latency issues, and potential mitigation strategies.

\subsection*{Inference Costs}

Using LLMs involves significant API costs. For example, the cost per API call can be calculated as:
\[
\text{Cost per call} = \left(\frac{\text{tokens per call}}{1000}\right) \times \alpha
\]
where \(\alpha\) is the cost per 1000 tokens. For a system with \(n\) users and \(m\) prompts, the total cost \(C\) is:
\[
C = n \times m \times \text{Cost per call}
\]

\noindent For example, if an average recommendation call uses 1500 tokens, the cost per call is $0.03$. For 610 users with 7 prompts each, the total cost would be $128.10$. To mitigate these costs, we can explore strategies such as token optimization, batch processing, and selective use of LLMs. Our code implements these strategies by limiting the maximum tokens per response and grouping prompts to minimize the number of API calls.

\subsection*{Latency}
Latency, the time taken to generate recommendations, affects user experience. Observed latencies ranged from a \textit{few to over ten seconds per call. }Mitigation strategies include asynchronous processing to handle multiple requests concurrently, caching mechanisms to store common recommendations, and using faster, distilled versions of LLMs. Our code addresses latency by measuring and optimizing the time taken for each API call and incorporating mechanisms to batch and streamline requests. Additionally, practical considerations such as using cloud services for scalability, combining LLMs with traditional collaborative filtering models, and regularly evaluating performance metrics ensure the system remains efficient and cost-effective. }

\vspace{1mm}
\noindent \textbf{Summary.}
\begin{tcolorbox}[colback=yellow!10, title=Answers to Research Questions for Experiment 1 and 2, colframe=yellow!40!black, colbacktitle=yellow!85!black]

\textbf{Answer to RQ1:} The incorporation of various goal-oriented prompts  impacts the accuracy of GPT-based models, with some prompts enhancing performance while others, like diversity-focused prompts, reducing it. However, GPT-based models generally lag behind CF baselines in zero-shot settings. 
\vspace{1mm}

\textbf{Answer to RQ2:} GPT-based recommendation systems show consistent personalization performance across multiple runs, with minor variability. This stability is comparable to CF baselines, indicating reliable behavior despite inherent randomness in individual runs. 
\vspace{1mm}

\textbf{Answer to RQ3:} GPT-based recommendation systems exhibit lower provider fairness compared to CF baselines. However, incorporating a \dq{Fair Recommender} system role in GPT-based models improves fairness and diversity, demonstrating controllability and potential for mitigation of item unfairness.

\vspace{1mm}
\textbf{Answer to RQ4:} GPT-based models tend to recommend newer movies and show a broader genre distribution compared to CF baselines, which prefer older, well-established films. Including explicit user ratings in prompts enhances personalization performance significantly.
\vspace{1mm}

\textbf{Answer to RQ5:} Different ICL strategies impact RecLLM quality and biases variably, with zero-shot learning generally providing better accuracy, while ICL strategies, especially with demographic information, offer improvements in certain contexts. Prompt design plays a crucial role in these outcomes.
\vspace{1mm}

\textbf{Answer to RQ6:} The economic and practical implications of using GPT-based models include significant inference costs and latency issues. Mitigation strategies like token optimization, batch processing, and using distilled models can help manage these challenges, ensuring efficient and cost-effective system performance.

\end{tcolorbox}

\bluegreen{\section{Conclusion and Future Directions}

\label{sec:conc}
This research provides a comprehensive analysis of biases of Recommender Systems using Large Language Models (RecLLMs), specifically focusing on ChatGPT-based systems. Our study highlights the distinct capabilities and biases of these systems compared to traditional Collaborative Filtering (CF) methods within the domain of movie and music recommendations. The experimental results emphasize the significant impact of \textit{prompt design} strategies on various aspects of recommendation quality, including accuracy, provider fairness, catalog coverage, diversity, temporal stability, genre dominance, and recency. 

In particular, we conducted two experiments using different datasets to evaluate the effectiveness of these strategies: classical top-$K$ recommendations and sequential in-context learning (ICL).

Key contributions and findings of our work include:
\begin{itemize}
    \item \textbf{Enhanced Fairness and Diversity through Role-Based Prompts:} Utilizing role-based prompts (e.g., ``act as a recommender'' or ``act as a fair recommender'') significantly enhances fairness and diversity in recommendations, effectively mitigating item unfairness.
    \item \textbf{Sequential In-Context Learning (ICL):} Incorporating sequential in-context learning prompts (ICL-1 and ICL-2) shows improvements in recommendation accuracy in certain contexts, particularly with demographic information. For instance, in the MovieLens dataset with age-group context, the Hit Rate (HR) for ICL-2 improves compared to zero-shot learning. However, ICL prompts do not consistently lead to better item fairness and coverage compared to zero-shot learning.
    \item \textbf{Improved Personalization with Ratings:} Incorporating user ratings into prompts improves the personalization of recommendations, highlighting the potential for fine-tuning prompts to better reflect user preferences.
    \item \textbf{Temporal Bias towards Recent Releases:} GPT-based models exhibit a strong tendency to recommend newer movies, particularly those released post-2000, contrasting with the older preferences of CF models. This suggests that GPT models can capture recent trends better, making them more temporally relevant.
    \item \textbf{Item Fairness and Diversity:} While GPT-based models generally exhibit lower item fairness compared to classical CF models, scenarios involving fair recommender roles show improved fairness and diversity, as indicated by higher entropy values.
    \item \textbf{Stability in Performance Metrics:} Despite inherent randomness, GPT-based models show consistent performance across multiple runs, as evidenced by low standard deviations in metrics such as NDCG. This stability is crucial for reliable recommendation systems.
\end{itemize}

Our research identifies several key areas for further exploration that can push forward the capabilities and effectiveness of recommendation systems leveraging large language models (RecLLMs).  Expanding research to include various content domains such as books, music, and e-commerce, along with conducting long-term user engagement studies, will provide deeper insights into this research. Firstly, refining prompt design and enhancing generalization across various datasets and tasks is a crucial step to improving recommendation accuracy and relevance. Additionally, implementing more advanced few-shot learning techniques, such as carefully selecting which examples to use and how to present them, can enhance the contextual relevance and accuracy of recommendations. These examples could be chosen from similar users (peer users) to make the system behave more like CFg models, thereby combining the strengths of both approaches.

Moreover, integrating GPT with collaborative filtering (CF) in a Retrieval-Augmented Generation (RAG) framework could leverage the strengths of both methodologies, leading to more robust and effective recommender systems. Future work should also address additional fairness dimensions, such as ensuring consumer-side fairness to prevent recommendations from disproportionately favoring certain user groups. Calibration, or aligning recommendations with user expectations to offer a balanced mix of familiar and novel items, is another critical aspect. Furthermore, ensuring counterfactual fairness, where recommendations remain fair even when user attributes or behaviors are altered, is also an essential and interesting future direction.

We explicitly acknowledge the limitations of our current work in using a single LLM based on ChatGPT. Future studies should explore other large language models (LLMs) and compare their performance to further understand their potential and limitations in diverse recommendation scenarios.}

\bibliographystyle{ACM-Reference-Format}
\bibliography{refs_tors2024}

\appendix

\end{document}